\documentclass[11pt]{article}
 
\usepackage{amsmath,amsthm}
\usepackage{amsthm}
\usepackage{amssymb}
\usepackage{amsfonts} 
\usepackage{color}

\usepackage[latin1]{inputenc}

\usepackage{graphicx}

\newcommand{\sect}[1]{\setcounter{equation}{0}\section{#1}}

\def\1{\'{\i}}

\begin{document}

\bigskip
 
\begin{center}

{\Large{\bf Superintegrability of the Fock-Darwin system}}

\bigskip 

{\sc E. Drigho-Filho$^a$, \c{S}. Kuru$^b$, J. Negro$^c$ and L.M. Nieto$^c$ }
\bigskip

\noindent
$^a$Departamento de Fisica, Universidade Estadual Paulista-UNESP, \\ 15054-000 S. Jose do Rio Preto,
SP, Brazil
%, e-mail: : elso@ibilce.unesp.br
\\
\noindent
$^b$Department of Physics, Faculty of Science, Ankara
University, 06100 Ankara, Turkey
\\ 
 \noindent
$^c$Departamento de F\'{\i}sica Te\'orica, At\'omica y
\'Optica, Universidad de Valladolid,\\  47011 Valladolid, Spain
%\\~~E-mail:   jnegro@fta.uva.es

\bigskip

\today

\end{center}

\bigskip

\begin{abstract}
\noindent 
The Fock-Darwin system is analysed from the point of view of its symmetry properties
in the quantum and classical frameworks. The quantum Fock-Darwin system is known to have two sets of ladder operators, a fact which guarantees its solvability.
We show that for rational values of the quotient of two relevant frequencies, this system is superintegrable, 
the quantum symmetries being responsible for the degeneracy of the energy levels. These symmetries are of higher order and close a polynomial algebra.  
In the classical case, the ladder operators are replaced by ladder functions
and the symmetries by constants of motion. We also prove that the rational classical system is 
superintegrable and its trajectories are closed. The constants of motion are also generators of symmetry transformations in the phase space that have been integrated for some special cases.
These transformations connect different trajectories with the same energy. 
The coherent states of the quantum superintegrable system are found and they reproduce the closed trajectories of the classical  one.
\end{abstract}

\bigskip 


 \medskip


%%%%%%%%%%%%%%%%%%%%%%%%%
\sect{Introduction}

In this work, we will revisit the Fock-Darwin (FD) system  
\cite{fock,darwin} with two main purposes: to examine in 
close detail its symmetries and 
its superintegrability character, and to give a complete picture of the system
in both, the quantum and classical frameworks.
The FD system consists in a charged particle moving in the plane and confined by a harmonic potential under an external uniform magnetic field. Here, we are not taking into account the spin splitting
in the magnetic field since this can be directly added at any stage.

The FD system has a number of applications in several fields. 
For example, it is used as  frequent  ingredient of  quantum dots. Due to the small size (of a few nanometers), when  the
discrete energy levels are filled with electrons, the quantum dot is called artificial atom, an entity
whose properties have been recently described. If there are more than one electron confined in the quantum dot,
the Coulomb interaction has to be taken into account. In this case, approximation
methods, like diagonalization of the Hamiltonian matrix or the constant interaction model
\cite{Hawrylak93,Kouwenhoven01,Chakraborty94,Chakraborty07,Peeters04,Chakraborty12}, are available.

In works dealing with quantum dots, the connection between `accidental degeneracy' and the symmetry group of a Hamiltonian
has attracted considerable attention \cite{Johnson00}. This connection was studied long time ago for a Hamiltonian describing a
particle in a central potential \cite{Hill40,Winternitz75,Quesne83,Moshinsky84,Quesne86}.
In general terms, a quantum system of $n$ degrees of freedom is called integrable if it has $n$ algebraically independent symmetry operators, including
the Hamiltonian, commuting with each other. When there are additional symmetry operators
so that we have the maximum set of $2n-1$ independent symmetry operators (not necessarily commuting), the system is called
superintegrable (or sometimes maximally superintegrable) \cite{Miller13,Miller11,Ballesteros16}.
In the classical context, the symmetries are replaced by constants of motion, and
the commutativity by the vanishing of Poisson brackets.
In these definitions it is assumed that the symmetries (or constants of motion)
are polynomials in the momenta.

In this paper, we address the characterization of the symmetries of the quantum FD system
in a simple and consistent way.
First of all, let us remember that the FD system has two limiting cases, the isotropic harmonic oscillator (HO) and
the Landau system,
which are well known to be superintegrable
systems, with second order symmetries leading to several sets of separable coordinates.
However, for the generic FD system the situation is not so evident and depends on
the ratio between two relevant frequencies, as we will see later.  
Only if this ratio is rational the system (called ``rational'' quantum FD system) will be
superintegrable. In this special case,
the symmetries are of higher order (greater than two), a fact that will not allow for additional 
separable coordinate systems \cite{Miller13}. 
As a consequence of the different symmetries of HO, Landau and the general FD system, the corresponding eigenvalues
have also different degeneracy properties:
for the Landau system there is an infinite degeneracy, in the HO each level has a finite degeneracy, and in the FD system
there may be no degeneracy at all or a special finite degeneracy, depending on the above mentioned frequency ratio. 

In the classical FD system, instead of symmetries we have to consider constants of motion. It turns out that 
the ``rational'' classical FD system is a superintegrable system
where the bounded orbits of the motion are closed.
In addition, the higher order constants of motion directly supply the equations of the trajectories and some of its properties. But also, these constants of motion are
symmetry generators that will be studied in detailed. In particular, we will obtain
finite symmetry transformations for a number of cases. The classical HO and Landau systems are
also included as limiting cases.
 
In order to see the relation between classical and quantum phenomena, it is important to study quantum coherent states. The coherent states associated to the the FD system have been studied under different
conditions. The first contributions were due to Feldman and Mank'o \cite{Manko69,Manko71,Feldman70}, and 
more recent application have been given in \cite{Sudiarta08,Peres09,Dehghani13}.  Another important area where coherent states of FD type systems have been considered is in paraxial optics,
where similar Hamiltonians are used to describe some optical waves (the so called 
Hermite-Gaussian and Laguerre-Gaussian modes) \cite{Chen10,Chen11}. As in the present work we 
study the classical and quantum symmetry properties of the FD system, we have considered that, for the sake of completeness, it is also relevant to compute the  coherent states in order to complement both points of view.

This paper is organized as follows: In Section~2, 
we solve the eigenvalue problem for the generic FD system giving the
eigenfunctions and eigenvalues in polar coordinates.
In Section~3, we discuss the spectrum degeneracy and the symmetries of three particular cases:  HO, Landau, and
``rational'' FD systems.
Section~4 is devoted to the analysis of the trajectories (determined by symmetries) and the motion 
(obtained by ladder functions) of the classical FD system. The finite symmetry transformations generated by the constants of motion in the phase space are also examined for these three special cases.
In Section 5, we consider the connection of the classical motions and
the quantum coherent states. Finally, the last Section  contains a summary of the original results and contributions
of the paper.

%%%%%%%%%%%%%%%%%%%%%%%%%%
\sect{The quantum Fock-Darwin Hamiltonian}

The FD system consists in a particle of mass $m$ 
and charge $e$ moving in a plane under a harmonic oscillator potential
of constant $k$ and subject to a constant magnetic field of intensity
$B$ perpendicular to the plane. Using the symmetric
gauge for the vector potential, 
\begin{equation}
{\bf A}= \left(-\frac{B}2\, y, \frac{B}2\, x, 0\right),\qquad
{\bf B} = (0, 0, B)\, ,
\label{potential}
\end{equation}
the quantum FD Hamiltonian is
\begin{equation}
\widetilde H=\frac1{2\,m}\left(P_x + \frac{e\, B}{2\,c} y\right)^2
+ \frac1{2\,m}\left( P_y - \frac{e\,B}{2\,c} x\right)^2
+\frac{k}2\left(x^2+y^2\right)\,,
\label{hq}
\end{equation}
where $c$ is the speed of light, and $P_x=-i\,\hbar\,\partial_x$, $P_y=-i\,\hbar\,\partial_y$  are momentum operators, 
with the following  notation $\partial_x=\partial/\partial x$ and $\partial_y=\partial/\partial y$.
The corresponding stationary Schr\"odinger equation describing this system 
in Cartesian coordinates is given by  
\begin{equation}
\left\{-\frac{\hbar^2}{2\,m}\left(\partial_x^2 + \partial_y^2 \right)
+\left(\frac{m\,\omega_c^2}8 + \frac{k}2\right)\left(x^2+y^2\right)
+i\, \frac{\hbar \,\omega_c}2 (x\,\partial_y - y\,\partial_x) \right\}\widetilde\Psi(x,y)
= E\,\widetilde\Psi(x,y)\,.
\label{h}
\end{equation}
Besides the Larmor (or cyclotron) frequency $\omega_c$,  there are
other relevant frequencies  involved here: 
\begin{equation}
\omega_c = \frac{e\, B}{m\,c},\qquad
\omega_o = \sqrt{\frac{k}{m}}\,,\qquad
\omega= \sqrt{\frac{\omega_c^2}4 + \omega_o^2} \, ,\qquad
\gamma = \frac{\omega_c/2}{\omega}
=\frac{eB}{\sqrt{e^2B^2+4mkc^2}} \, ,
\label{coeff}
\end{equation}
 where $\omega_o$ is the natural frequency of the oscillator, 
$\omega$ is a FD characteristic frequency and  
$\gamma$ is a ratio of frequencies.

Since this system has a geometric rotational symmetry around the
$z$-axis, it is convenient to write the  Hamiltonian  \eqref{h} in polar coordinates
$(r, \varphi)$ and at the same time to change the expression for the eigenfunction as
\begin{equation}
\widetilde\Psi(x,y) = r^{-1/2}{ \Psi}(r, \varphi) \,.
\label{newpsi}
\end{equation}
It is also convenient to express the eigenvalue equation in terms of the dimensionless variable $\rho$ and  parameter $\varepsilon$, defined as follows:
\begin{equation}
\rho = \sqrt{\frac{m\,\omega}{\hbar}}\, r,\qquad
\varepsilon 
= \frac{E}{\hbar\, \omega}\, .
\label{param}
\end{equation}
Then, from \eqref{h} the corresponding eigenvalue equation takes the form
\begin{equation}
 H  \Psi (\rho, \varphi)= \frac12\left[-\partial_\rho^2
- \frac{1/4 +\partial_\varphi^2}{\rho^2} +\rho^2 
+2 i\,\gamma\partial_\varphi \right] \Psi(\rho, \varphi) = \varepsilon\,  \Psi (\rho, \varphi)\,.
\label{hrho}
\end{equation}
In the sequel, we  will allow for both positive and negative values of the Larmor frequency $\omega_c$
in order to take into account the two possible signs of the product $e\,B$. Observe that 
$-1\leq \gamma \leq 1$, $\gamma$ taking the values  $\pm1$ for a pure Landau system
and  zero for a pure HO. 
From (\ref{hrho}) we see that, apart from the dimensionless energy $\varepsilon$, the only parameter remaining in
the FD equation is just the coefficient $\gamma$, which  plays a key role in this system.

%%%%%%%%%%%%%%%%%%%%%%%%%%%%%%
\subsection{Quantum algebraic treatment}

As we have foreseen, the Hamiltonian (\ref{hrho}) explicitly commutes with
the angular momentum operator $ \widetilde L=-i \hbar \partial_\varphi$, which due to the  units used in the equation will be replaced by $ L=-i \partial_\varphi$.
Hence, we can look for separated solutions
\begin{equation}
 \Psi(\rho,\varphi) = R(\rho) \Phi(\varphi) \,.
\label{sep}
\end{equation}
The angular part of wave function must take the form
\begin{equation}
\Phi_\ell(\varphi)  = e^{i\, \ell \, \varphi},
\qquad 
 L\Phi_\ell(\varphi) = \ell\, \Phi_\ell(\varphi) \,,
\end{equation}
where, in order to have a single valued function, the  parameter $\ell$
must be restricted to integer values: $\ell = 0, \pm 1,\pm 2\dots$  
 The radial part $R(\rho)$ in (\ref{sep}) must be a square integrable
solution of
the reduced one-dimensional problem 
\begin{equation}
{H}_\ell R(\rho) = \frac12\left[-\partial_\rho^2
+ \frac{\ell^2-1/4 }{\rho^2} +\rho^2 
-2 \gamma\,\ell \right]R(\rho) = \varepsilon\, R(\rho)\,.
\label{eqr}
\end{equation}
A similar equation in the variable $\rho$ is well known to appear when the factorization
method is applied to the radial oscillator \cite{Drigo}, except for the presence of the additional term with $\gamma$ coefficient. 
It has been shown in previous references \cite{FNO} that 
this radial Hamiltonian can be factorized in 
two ways by means of two sets of differential operators
\begin{eqnarray}
a_\ell^\pm =  
\frac1{2}\left(\mp\partial_\rho - \frac{\ell+1/2}{\rho} +\rho\right)\,,
\qquad
b_\ell^\pm =  
\frac1{2}\left(\mp\partial_\rho + \frac{\ell-1/2}{\rho} +\rho\right)\,,
\label{abpm}
\end{eqnarray} 
as follows:
\begin{equation}
{H}_\ell = 2 a_\ell^+ a_\ell^- + \ell(1-\gamma)+1 
= 2 b_\ell^+ b_\ell^- - \ell(1+\gamma)+1 \,.
\label{habl}
\end{equation}
These two formulas lead to another expression for ${ H}_\ell$ in terms
of the $a_\ell^\pm,\,b_\ell^\pm$ operators (excluding $\ell$) \cite{kikoin}:
\begin{equation}
{H}_\ell = (1+\gamma)a_\ell^+ a_\ell^- 
+ (1-\gamma)b_\ell^+ b_\ell^- +1 \,.
\label{hab}
\end{equation}

All the previous relationships for radial operators $a_\ell^\pm, \, b_\ell^\pm$ 
can be translated, with some care,
into relations for ``dressed'' operators $a^\pm,\,b^\pm$ 
in both polar coordinates, defined as
\begin{eqnarray}
&&a^- =  
\frac1{2}e^{i\,\varphi}\left(\partial_\rho - \frac{-i\partial_\varphi+1/2}{\rho} +\rho\right)\,,\qquad \ a^+ = (a^-)^\dagger\,, 
\label{abpma}
\\[2.ex]
&&b^- =  
\frac1{2}e^{-i\,\varphi}
\left(\partial_\rho - \frac{i\partial_\varphi+1/2}{\rho} +\rho\right)\,,\qquad 
\ b^+ = (b^-)^\dagger\,.
\label{abpmb}
\end{eqnarray} 
Using the dressed operators, the factorization properties (\ref{habl}) come into
\begin{equation}
{H} = 2\, a^+ a^- + (1-\gamma) L + 1 
= 2 \,b^+ b^- -(1+\gamma) L + 1 \,.
\label{hablb}
\end{equation}
From these relations we get the following expressions for the angular momentum $L$
\begin{equation}\label{l}
 L = b^+b^- - a^+ a^- \,,
\end{equation}
and for the FD Hamiltonian
\begin{equation}
{ H} = (1+\gamma)a^+ a^- 
+ (1-\gamma)b^+ b^- +1 \,.
\label{habb}
\end{equation}
It is easy to prove that the  operators $a^\pm$ and $b^\pm$ constitute two independent realizations of the Heisenberg algebra:
\begin{equation}
[a^-,a^+] = 1,\qquad [b^-,b^+] = 1,\qquad [a^\pm,b^\pm] = 0\, .
\end{equation}
The corresponding number operators are given by $M=a^+a^-$ and $N=b^+b^-$. 
Taking into account the expression (\ref{l}), it is also immediate
to check that
\begin{equation}
[a^\pm,  L] = \pm\,a^\pm,\qquad
[b^\pm,  L] = \mp\,b^\pm \,.
\label{lpm}
\end{equation}
In other words, $a^+$ and $a^-$ acting on eigenfunctions of $L$
decreases and increases, respectively,  the eigenvalue in one unit; the action of $b^\pm$
have the opposite effect.

%%%%%%%%%%%%%%%
\subsection{Eigenfunctions and energies}

By means of the above algebraic properties,  
the FD Hamiltonian (\ref{habb}) can be written in terms of the number operators as
\begin{equation}\label{habb2}
{H}= (1+\gamma)M 
+ (1-\gamma)N +1 \, .
\end{equation} 
The eigenfunctions of (\ref{habb2}) will be labeled by two positive integer numbers,
$m,n=0,1,2,\dots$, corresponding to the number operators $M$ and $N$, respectively, and are given by the action of the creation operators on a fundamental eigenfunction $\Psi_{0,0}(\rho, \varphi)$:
\begin{equation} \label{mn}
 \Psi_{m,n} (\rho, \varphi)= \frac1{\sqrt{m!n!}} (a^+)^m(b^+)^n \Psi_{0,0} (\rho, \varphi)\, .
\end{equation}
The ground state wavefunction is determined by the conditions
\begin{equation}
a^-  \Psi_{0,0}(\rho, \varphi) = b^-   \Psi_{0,0}(\rho, \varphi)= 0
\ \implies\   \Psi_{0,0}(\rho,\varphi) = K_0\,\rho^{1/2} e^{-\frac12 \rho^2}\,,
\end{equation}
where $K_0$ is a normalization constant.

According to (\ref{habb2}) and (\ref{param}), the eigenvalues corresponding to these eigenfunctions are
\begin{equation}
\varepsilon_{m,n} = m(1+\gamma) + n(1-\gamma) +1 
\ \implies\  E_{m,n} = \big[m(1+\gamma) + n(1-\gamma) +1\big]\hbar\omega \, .
\label{emn}
\end{equation}
Therefore, the action of $a^+$ on an eigenfunction of $H$ produces another one with eigenenergy $E$ increased in $(1+\gamma)\hbar \omega$. On the other hand, the operator $b^+$ has a similar action on the eigenfunctions, but with jumps of $(1-\gamma)\hbar \omega$ units. Since $-1\leq\gamma\leq 1$, 
the contribution of these creation operators to the total energy is different, in general. We say that $a^\pm$ and $b^\pm$ are ladder
operators of the FD system with ``different steps''. It is well known that in quantum mechanics ladder operators are quite helpful to
determine the spectrum of a Hamiltonian, as we have just seen, but 
their classical counterpart may also play a relevant role in order to find the classical trajectories
\cite{Kuru}. The connection between classical
ladder functions and quantum ladder operators constitutes a general basis
to construct coherent states \cite{Nieto06,Cruz08,Campoamor12}, as it will be illustrated later
 in Section~5.

Coming back to the eigenfunctions \eqref{mn} of the Hamiltonian (\ref{habb2}), and taking into account \eqref{l}, they are also eigenfunctions of the angular momentum:
\begin{equation}
 L  \Psi_{m,n} (\rho, \varphi)= (n-m) \Psi_{m,n} (\rho, \varphi)\ \implies\ 
\ell= n-m \, ,
\label{eigenl}
\end{equation}
a result that is consistent with the commutation rules (\ref{lpm}).
In order to make explicit the dependence on the angular momentum, one can change the notation of these eigenfunctions as follows: from (\ref{eigenl}), the different eigenfunctions
corresponding to the same eigenvalue $\ell$ will be denoted by 
$ \Psi^\ell_p(\rho, \varphi)$,   $p=0,1,2\dots$, and can
be expressed in terms of $\Psi_{m,n}(\rho, \varphi)$, as 
\begin{equation}
 \Psi^\ell_p (\rho, \varphi)= 
\left\{
\begin{array}{ll} \Psi_{p,p+|\ell|}(\rho, \varphi),\ & {\rm for}\ \ell \geq 0,
\\[1.ex]
 \Psi_{p+|\ell|,p}(\rho, \varphi),\ &{\rm for}\  \ell \leq 0,
\end{array}\right.\qquad p=0,1,2,\dots;\ \ell=0,\pm1,\pm2,\dots\,
\end{equation}
Then, replacing this definition in (\ref{emn}), the eigenvalues are given by
\begin{equation}
E^\ell_p = (2p + |\ell| +1 )\,\hbar \omega - \frac12 \hbar\,\ell \,\omega_c\,.
\end{equation}
The corresponding eigenfunctions can be obtained by the action of the ladder operators, as in (\ref{mn}). For instance,
for $n\geq m$ (or $\ell\geq0$), we have
\begin{equation}
 \Psi^{\ell}_p(\rho,\varphi) = \frac1{\sqrt{(\ell+p)!p!}}(a^+)^p(b^+)^{\ell + p}\Psi_{0,0}(\rho, \varphi)\, ,
\end{equation}
and after straightforward computations we get
\begin{equation} \label{psiL}
 \Psi^{|\ell|}_p(\rho,\varphi) =  K^{|\ell|}_p\,e^{i\,\ell\,\varphi}\,\rho^{|\ell|+1/2}\,e^{-\rho^2/2}\,L_p^{|\ell|}(\rho^2) \,,
\end{equation}
where $L_p^{|\ell|}(\rho^2)$ are Laguerre polynomials and $K^{|\ell|}_p$ are normalization constants. The notation in terms of $(\ell,p)$ has been used in some previous references, but hereafter we adopt $\Psi_{m,n}(\rho, \varphi)$ with $(m,n)$ labeling  the eigenfunctions.

%%%%%%%%%%%%%%%%%%%%%%%%%%%%%%%%%%%%
\sect{Particular cases of the superintegrable quantum FD system}

In this section we will consider  three particular cases of the 
FD system: the isotropic harmonic oscillator, the Landau system
and, specially, the superintegrable ``rational'' FD system.

%%%%%%%%%%%%%%%%%%%%%%%%%%%%%%%%%%%%
\subsection{The two-dimensional isotropic harmonic oscillator}

First of all, we  will recall some known results for the two dimensional
isotropic harmonic oscillator (HO) that can be obtained as a limit of the FD system
when the magnetic field $B$ goes to zero.
In this case, the relevant magnitudes \eqref{coeff}-\eqref{param} take the specific values
\begin{equation}
\omega=  \omega_o \, ,\qquad
\omega_c=\gamma = 0,\qquad
\rho = \sqrt{\frac{m\,\omega_o}{\hbar}}\, r,\qquad
\varepsilon = \frac{E}{\hbar\, \omega_o}\, ,
\label{coeffho}
\end{equation}
the Hamiltonian (\ref{habb}) has the special form
\begin{equation}
H^{HO} =  a^+ a^- 
+  b^+ b^- +1=M+N+1\,,
\label{habbHO}
\end{equation}
and the eigenvalues (\ref{emn}) in this case are
\begin{equation}
\varepsilon_{m,n} = m + n +1 
\ \implies\  E_{m,n} = (m + n +1) \hbar\omega_o \, \qquad m,n= 0,1,2,\dots ,
\label{emnHO}
\end{equation}
which correspond to the sum of the spectra of two independent one-dimensional harmonic oscillators.
Now, let us analyze in some detail the symmetries and degeneracy of the Hamiltonian \eqref{habbHO}. As can be
seen from (\ref{emnHO}), the energy levels are degenerate: the
states $\Psi_{m,n}(\rho, \varphi)$ and $\Psi_{m',n'}(\rho, \varphi)$ have 
the same energy when $m+n=m'+n'=k\in\{ 0,1,2\dots\}$. If
this energy level is labeled by $E^{\rm H O}_k\equiv E_{m,n} = E_{m',n'}$,
then its degeneracy is $k+1$.

A set of symmetry operators for this Hamiltonian $H^{HO}$ is easily obtained from 
\eqref{habbHO}:
\begin{equation}
M= a^+a^-\,,\qquad N =  b^+b^-\,, 
\qquad S^- = a^+b^-,\qquad S^+ = a^-b^+\,.
\label{symHO}
\end{equation}
The symmetries $M$ and $N$ fix an eigenfunction state $\Psi_{m,n}$,
and the action of the other symmetries $S^\pm$ on this state will give us
the whole degeneracy subspace of constant energy. 
We should remark that the angular momentum operator $ L$ is
another symmetry, since according to (\ref{l}) $ L = N-M$. 
In total, there are three independent symmetry operators, for instance
$M,\, N, \,S^+$, and therefore we can say that this system is  superintegrable.
Although  the four symmetries (\ref{symHO})
are functionally dependent, they are useful to construct the $u(2)$ symmetry algebra. Indeed, if we introduce the following basis
\begin{equation}
\left\{ S=\frac12(N-M) = \frac12  L,\ S^\pm,\  H= N+M+ 1 \right\},
\end{equation}
we get a realization of $u(2)$ in the form $u(2) = su(2) \oplus u(1) = \langle S\,, S^\pm\rangle \oplus \langle  H \rangle$:
\begin{eqnarray}
[S,S^\pm] = \pm S^\pm,\qquad
[S^-,S^+]= -2 S,\qquad
[{ H}, \cdot \ ] = 0\,.
\label{hoalgebra}
\end{eqnarray} 
We should remark the well known fact that as the superintegrability is realized by second order symmetries, this system is separable in more than one set of coordinates: Cartesian, polar and 
elliptic \cite{Miller13}.

%%%%%%%%%%%%%%%
\subsection{The Landau system}

If the harmonic oscillator term is null ($k=0$) in the FD system,
we just have a charged particle in a constant magnetic field, 
which is the so-called Landau system. The characteristic values of the parameters
(\ref{coeff})-(\ref{param}) for this case are
\begin{equation}
\omega_o = 0\,,\qquad
\omega= \frac{\omega_c}2   \, ,\qquad
\gamma = \pm1,\qquad
\rho = \sqrt{\frac{m\,\omega_c}{2\hbar}}\, r,\qquad
\varepsilon = \frac{2E}{\hbar\, \omega_c}\, .
\label{coeffL}
\end{equation}
The sign of $\gamma$ depends on the sign of the product of the charge
and the field, $eB$. In what follows we adopt the positive sign, $\gamma=1$,
but equivalent considerations apply to the negative one.
Here, the Hamiltonian (\ref{habb}) takes the special form
\begin{equation}
 H^{\rm L} =  2\,a^+ a^-  +1=2\,M+1\,,
\label{habbL}
\end{equation}
and the eigenvalues (\ref{emn}) are independent of $n$:
\begin{equation}
\varepsilon^{\rm L}_{m} = 2\,m  +1 
\ \implies\  E^{\rm L}_{m} = \big(m  +1/2\big)\hbar\omega_c \,,\qquad
m=0,1,2\dots 
\label{emnL}
\end{equation}
In this case, we can appreciate that the energy levels have
infinite degeneracy because the energy does not depend on the second
quantum number $n$: the states $ \Psi_{m,n}(\rho, \varphi)$, with
the same $m$-value have the same energy, that
we called $E^{\rm L}_m$. The values of $E^{\rm L}_m$ are the half odd
multiples of the step energy $\hbar\, \omega_c$, where $\omega_c$
is the cyclotron frequency, corresponding to the spectrum of a one-dimensional HO.

For Landau system,
we have the following symmetry operators:
\begin{equation}
M= a^+a^-\,,\qquad N =  b^+b^-\,,\qquad  S^- = b^-\,,\qquad S^+ = b^+\,,
\label{symL}
\end{equation}
which allow us to describe the degeneracy of the system: 
each state $ \Psi_{m,n}(\rho, \varphi)$ is characterized by the symmetries $M$ and $N$, while the
other symmetries $S^\pm=b^\pm$ acting on that state generate the whole  subspace of energy $E^{\rm L}_m$.
As in the HO limit, $ L=N-M$ is another symmetry operator.
There are three independent symmetry operators (for instance $M, b^\pm$),
so that this system is also superintegrable. In this case, we can identify
the symmetry Lie algebra as $os(1)\oplus u(1)$, where $os(1) = \langle N, S^\pm \rangle$ is the
one dimensional oscillator algebra, and $u(1) = \langle  H \rangle$  commutes
with the other generators:
\begin{eqnarray}
[ H,\cdot\ ]=0,\qquad[S^-,S^+]=1,\qquad [N,S^\pm]=\pm S^\pm \,.
\label{landaualgebra}
\end{eqnarray} 

 If we had chosen the other sign for
$\gamma$, the roles of the operators $a^\pm$ and $b^\pm$ would
have been exchanged in the previous discussion. As the maximum order of the symmetries (\ref{symL}) is two, there are more than one separable coordinate set: Cartesian and polar.

%%%%%%%%%%%%%%%
\subsection{Quantum superintegrable rational FD system}

\begin{figure}[htb]
\centering
\includegraphics[width=0.45\textwidth]{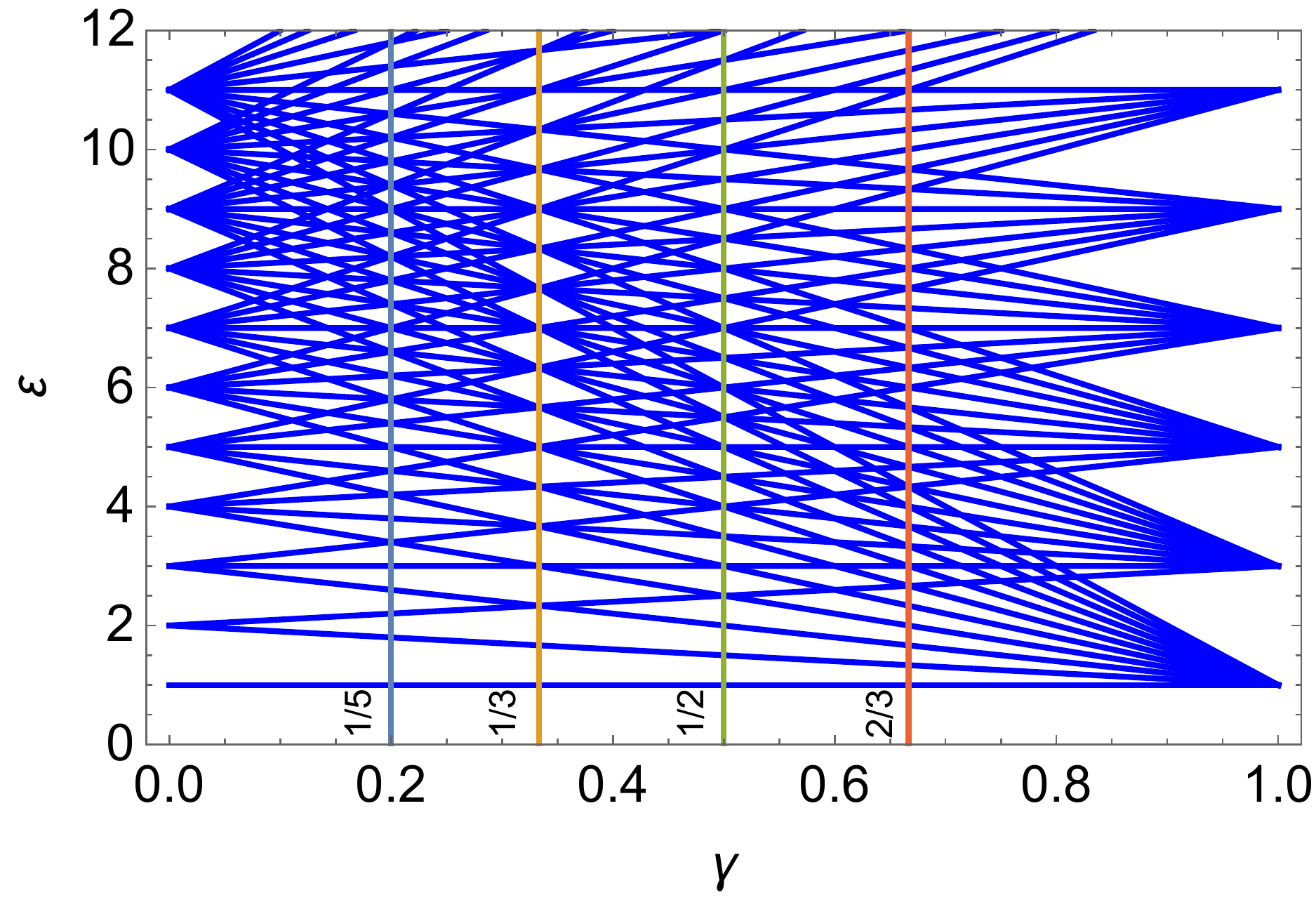}
\qquad 
\includegraphics[width=0.44\textwidth]{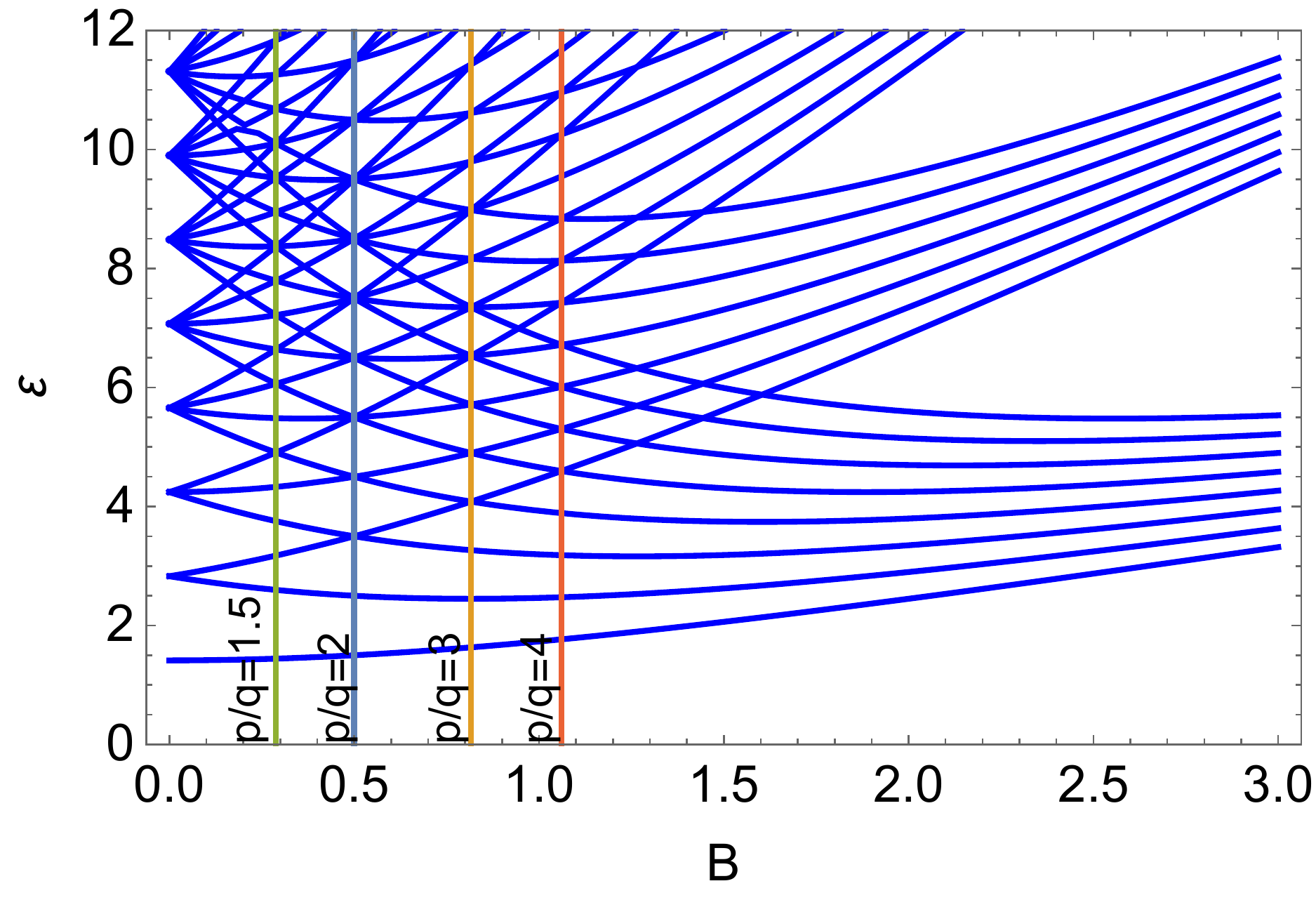}
\caption{\small Plot of the dimensionless energy levels $\varepsilon_{m,n}$ for FD system \eqref{emn}: on the left as a function of $\gamma$; on the right depending on
magnetic field $B$ (with $mkc^2/e^2=1$). The vertical lines correspond 
to rational values of $(1+\gamma)/(1-\gamma)=p/q$ (right) and $\gamma$ (left), where
the system is superintegrable and the energy levels have finite degeneracy.
\label{niveles}}
\end{figure}

We have seen  in the previous subsections that for the special values $\gamma=0$ and $\gamma=\pm1$, corresponding to the HO and Landau  limits for \eqref{habb2}, we obtain superintegrable systems.  Then, a natural question arises: Is the FD system
 also superintegrable for other values of $\gamma$ such that $0<|\gamma|<1$? To answer this query, 
let us assume that the coefficient $\gamma$ is a rational number, in which case we will write
\begin{equation}
\frac{1+\gamma}{1-\gamma} = \frac{p}{q},\qquad
p,q\in \mathbb N\,,
\end{equation}
where $p,q$ have no common non-trivial integer factors.
Then, it is easy to show that the FD system admits the following symmetry operators:
\begin{equation}
M= a^+a^-\,,\qquad N=  b^+b^-\,, 
\qquad S^- = (a^+)^q(b^-)^p,\qquad S^+ = (a^-)^q(b^+)^p\,.
\label{symFD}
\end{equation}
As in the previous cases, $M$ and $N$ determine an eigenfunction $\Psi_{m,n}(\rho, \varphi)$ and the symmetries $S^\pm$ applied to this state  produce all the remaining degenerate eigenstates.
The description of the eigenspaces is similar, with some peculiarities that we will comment next.
First of all, let us characterize the nondegenerate energy levels (eigenspaces of dimension one): there are $p\times q$ such eigenspaces, each one spanned by one of the eigenfunctions:
\begin{equation}
 \Psi_{m,n} (\rho, \varphi), \qquad {\rm with}\qquad 0\leq m<q,\quad  0\leq n<p\,.
\end{equation} 
Now, let us consider the degenerate energy levels of dimension $k+1$. They are spanned by the eigengunctions
\begin{equation}
 \Psi_{k_1q+m,k_2p+n} (\rho, \varphi), \quad {\rm with}\quad 0\leq m<q,\ 0\leq n<p\, \quad {\rm and}\quad
k_1+k_2 = k,
\end{equation} 
where $k,m,n$ are fixed and $k_1,k_2$ take the values $0,1,2\dots$ The eigenfunctions $\Psi_{k_1q+m,k_2p+n}(\rho, \varphi)$
are connected among them by the $S^\pm$ symmetries.
In total, there are $p\times q$ eigenspaces with the same
degeneracy dimension. This degeneracy property can be seen in Figure~\ref{niveles},  where
the energy levels in \eqref{emn}, $\varepsilon_{m,n} = m(1+\gamma) + n(1-\gamma) +1$, are plotted as a function of either $\gamma\in[0,1]$ (left) or the magnetic field (right).
We observe that: 
\begin{itemize}
\item
For $B=0$, then $\gamma=0$, and we recover the spectrum of the HO
with finite  degeneracy, given in \eqref{emnHO}. 
\item
When $B\to +\infty$, then
 $\gamma\to 1$, and we get the Landau levels, where there is an infinite degeneracy. 
\item
When $B$ is such that $(1+\gamma)/(1-\gamma)= p/q$ is rational, we have also a superintegrable FD system with 
nontrivial symmetries $S^\pm$, giving rise to a finite degeneracy of the energy levels (see some especific values in Figure~\ref{niveles}). The number of eigenspaces with the same degeneracy dimension is $p\times q$.
\end{itemize}

%%%%%%%%%%%%%
%%%%%%%%%%%%%

Notice that $ L=N-M$ is also a symmetry operator.
As in the previous cases, we
have three independent symmetry operators and we conclude that for
rational values of $\gamma$ the system is superintegrable; we will
call it ``the rational FD system''.
In fact, we can see that the symmetry operators (\ref{symHO}) and (\ref{symL}),
correspond to  special cases of (\ref{symFD}): for $p=q=1$ (HO) and $p=1, \,q=0$ (Landau).

However, there are differences in this rational case with respect to the
HO and Landau systems which are worth to comment.
The first one is that the symmetry operators $S^\pm$ are of $p+q$ order, always higher
than two. This means that they do not produce other
separable set of coordinates besides the polar coordinates \cite{Miller13}.

The second difference is that the symmetry operators given by (\ref{symFD}) satisfy a polynomial algebra (not a Lie algebra):
\begin{eqnarray}
&&[M,N]=0,\qquad[S^-,S^+]=P_1(M,N)-P_2(M,N),\nonumber \\[1.ex]
&&[M,S^{\pm}]=\mp q\,S^{\pm},\qquad [N,S^{\pm}]=\pm p\,S^{\pm},
\label{fdalgebra}
\end{eqnarray} 
where $P_1(M,N)$ and $P_2(M,N)$ are the following polynomials of orden $p+q$ on $M,\,N$:
\begin{eqnarray*}
&&P_1(M,N)=S^-\,S^+=M(M-1)\dots (M-q+1)\,(N+1)\dots (N+p)\,,\nonumber \\[1.ex]
&&P_2(M,N)=S^+\,S^-=(M+1)\dots (M+q)\,N(N-1)\dots (N-p+1)\,.
\end{eqnarray*} 
Hence,  $[S^-,S^+]$ is a polynomial in $M$ and $N$  of degree $p+q-1$. For $p=q=1$ and $p=1, \,q=0$, we recover the symmetry algebras of  the HO and Landau systems, respectively.

%%%%%%%%%%%%%%%%%%%%%
\sect{The classical FD system}

In this section, we will study the motion and trajectories of the classical FD system, a task that will be done using ladder functions
$\alpha^\pm, \beta^\pm$, associated to the quantum ladder operators $a^\pm,b^\pm$ 
defined in Section 2.

We start with the classical Hamiltonian $\tilde h$  corresponding to
the quantum Hamiltonian $\widetilde H$ of (\ref{hq}), where
the momentum operators $P_x,\, P_y$ and position operators
$x,y$ have been replaced by their canonical variables.
Next, we perform a change from Cartesian to polar coordinates $(r,\varphi)$,
and then, to dimensionless radial coordinates $\rho,\,p_\rho$ given
by
\begin{equation}
\rho = \sqrt{m\,\omega}\, r,\qquad
p_\rho = \frac1{\sqrt{m\,\omega}}\, p_r,
\end{equation}
in agreement with the quantum counterpart (\ref{param}). Finally, the 
reduced classical Hamiltonian $\tilde h/\omega=  h$ takes the form
\begin{equation}
 h  \equiv h(\rho,p_\rho,\varphi,p_\varphi )= \frac12 \left[p_\rho^2
+ \frac{p_\varphi^2}{\rho^2} +\rho^2 
-2 \,\gamma\, p_\varphi \right] \,,
\label{hrhoC}
\end{equation}
where we  are using the definitions (\ref{coeff}) for the frequencies 
$\omega_c,\,\omega_o,\,\omega$ and the coefficient $\gamma$. As this system has rotational symmetry,  $h$ 
does not depend on $\varphi$ and the angular momentum $p_\varphi$ is a constant of motion,
$p_\varphi=\ell\in \mathbb R$.  Hence, the 
effective Hamiltonian obtained from (\ref{hrhoC}) is 
\begin{equation}
 h_{\rm eff}   \equiv h_{\rm eff}(\rho,p_\rho)= \frac12 (p_\rho^2+ V_{\rm eff}(\rho)),\qquad
V_{\rm eff}(\rho)=\frac{\ell^2}{\rho^2} +\rho^2 
-2 \,\gamma\, \ell\,,
\label{veff}
\end{equation}
where $V_{\rm eff}(\rho)$ is an effective potential.

%%%%%%%%%%%%%%%%%%%%%%%%%%%%%%
\subsection{Classical algebraic treatment}

Now, we define the classical analogs of the ladder operators (\ref{abpma})-(\ref{abpmb})
as
\begin{eqnarray}
\alpha^\pm =  
\frac1{2}e^{\mp i\,\varphi}\left(\mp i p_\rho - \frac{p_\varphi}{\rho} +\rho \right)\,,
\qquad
\beta^\pm =  
\frac1{2}e^{\pm i\,\varphi}
\left(\mp i p_\rho + \frac{p_\varphi}{\rho} +\rho\right)\,.
\label{bpmbC}
\end{eqnarray} 
Notice that $\alpha^+$, $\beta^+$ are the complex conjugate functions of 
$\alpha^-$, $\beta^-$, respectively.
Indeed, these functions satisfy the Poisson bracket relations of the direct sum of two classical Heisenberg 
algebras:
\begin{equation}
\{\alpha^-,\alpha^+\}=-i,\qquad
\{\beta^-,\beta^+\}=-i,\qquad
\{\alpha^\pm,\beta^\pm\}=0\,,
\end{equation}
where $\{\cdot,\cdot\}$ are Poisson brackets:
$$ 
\{ f,g \}= \frac{\partial f}{\partial \rho}\frac{\partial g}{\partial p_\rho}-\frac{\partial f}{\partial p_\rho}\frac{\partial g}{\partial \rho} +
\frac{\partial f}{\partial \varphi}\frac{\partial g}{\partial p_\varphi}-\frac{\partial f}{\partial p_\varphi}\frac{\partial g}{\partial \varphi} .
$$
The classical Hamiltonian $h$ can be expressed in a  form that resembles the quantum case  \eqref{habb}:
\begin{eqnarray}
h = 2\alpha^+ \alpha^- 
+ (1-\gamma)\ell = 2 \beta^+ \beta^- -(1+\gamma)\ell 
 = (1+\gamma)\alpha^+ \alpha^- 
+ (1-\gamma)\beta^+ \beta^- \,.
\label{habbC}
\end{eqnarray}

%%%%%%%%%%%%%%%%%%%%%%%%%
\subsection{Constants of motion and trajectories of the classical FD system}

If $\gamma$ is a rational number then $(1+\gamma)/(1-\gamma) = p/q$ is rational too, 
the  classical FD system is superintegrable and has the following constants of motion, which are the analogs of the quantum symmetry operators for the quantum superintegrable FD system:
\begin{equation}
{\cal M}= \alpha^+\alpha^-\,,\qquad {\cal N}=  \beta^+\beta^-\,, 
\qquad {\cal S}^- = (\alpha^+)^q(\beta^-)^p,\qquad {\cal S}^+ = (\alpha^-)^q(\beta^+)^p\,.
\label{symFDclassical}
\end{equation}
The constants of motion ${\cal S}^+$ and ${\cal S}^-$ are complex conjugate of each
other. The angular momentum is also a constant of motion that can be expressed as 
\begin{equation}{\cal L}={\cal N}-{\cal M}\,.%= |\beta^+(0)|^2-|\alpha^+(0)|^2\,.
\label{symFDclassicalzz}
\end{equation} 
Remark that from (\ref{bpmbC}) one can check that ${\cal L} = p_\varphi$.
The constants of motion (\ref{symFDclassical}) satisfy the following Poisson brackets:
\begin{eqnarray}
&&\{{\cal M},{\cal N}\}=0,\qquad \{{\cal S}^-,{\cal S}^+\}=i\,q^2\,{\cal M}^{q-1}\,{\cal N}^p -i\,p^2\,{\cal M}^q\,{\cal N}^{p-1},\nonumber \\[2.ex]
&&\{{\cal M},{\cal S}^{\pm}\}=\pm \,i\,q\,{\cal S}^{\pm},\qquad \{{\cal N},{\cal S}^{\pm}\}=\mp \,i\,p\,{\cal S}^{\pm}\,.
\label{fdalgebraclassical}
\end{eqnarray} 
It can be shown that these classical Poisson brackets are the limit of the quantum commutators given in (\ref{fdalgebra})
according to the Dirac quantization rule: 
$[\hat A, \hat B] = i \hbar\, \hat C \implies   \{A,B\} = C$.

Let us emphasize that there are only three independent constants of motion. For example,
the set $\{ h, p_\varphi, {\cal S}^+\}$ is a possible choice. We can check that 
\begin{equation}
{\cal S}^+ {\cal S}^- = |{\cal S}^\pm|^2 =
2^{-(p+q)}\big( h +(\gamma-1)p_\varphi\big)^q
\big( h +(\gamma+1)p_\varphi\big)^p\, .
\end{equation}
Notice that due to \eqref{habbC}
we have the following inequalities:
\begin{equation}
\varepsilon +(\gamma-1)\ell\geq0\,,\qquad
\varepsilon +(\gamma+1)\ell\geq 0 \,.
\end{equation}
Therefore, ${\cal S}^+$ can be written as
\begin{equation}\label{sp1}
{\cal S}^+ =e^{i\varphi_0} |{\cal S}^+| =e^{i\varphi_0}\,  \frac1{2^{(p+q)/2}}\big( \varepsilon+(\gamma-1)\ell\big)^{q/2}
\big(\varepsilon +(\gamma+1)\ell\big)^{p/2}\,,
\end{equation}
where $|{\cal S}^+|$ depends only on $\varepsilon$ and $\ell$, while the phase $\varphi_0$ is the value characterizing the third constant of motion ${\cal S}^+$. On the other hand,  from \eqref{symFDclassical} and \eqref{bpmbC}, ${\cal S}^+$
can be expressed in terms of $\varphi$ and $\rho$  as follows 
\begin{equation}\label{sp2}
{\cal S}^+ = 
e^{i(p+q)\varphi}\frac{1}{2^{q+p}}\left( i p_\rho - \frac{p_\varphi}{\rho} +\rho\right)^q \left(- i p_\rho +\frac{p_\varphi}{\rho} +\rho\right)^p\,.
\end{equation}
 Hence, from (\ref{sp1})-(\ref{sp2}) and taking into account (\ref{hrhoC}), we find the
equation of the orbits depending on the three constants of motion $h=\varepsilon$, $p_\varphi=\ell$  and $\varphi_0$:
\begin{equation}\label{sp2new} 
 \frac{e^{i\varphi_0}}{2^{(p+q)/2}}\big( \varepsilon+(\gamma-1)\ell\big)^{q/2}
\big(\varepsilon +(\gamma+1)\ell\big)^{p/2} = \frac{e^{i(p+q)\varphi}}{2^{q+p}}\left( i p_\rho - \frac{p_\varphi}{\rho} +\rho\right)^q \left(- i p_\rho +\frac{p_\varphi}{\rho} +\rho\right)^p  .
\end{equation}
This equation shows that the trajectories are  $\varphi$-periodic  with
fundamental period $T= {2\pi}/(p+q)$. This important property is due to the 
existence of the third independent constant of motion ${\cal S}^+$ 
(or equivalently ${\cal S}^-$). By applying the previous formula, we will explicitly write the trajectories in polar coordinates for the
particular cases of HO and Landau: 
 
\begin{itemize}
\item
Trajectories of the HO system: $p=q=1$.
The corresponding equation becomes
\begin{equation}
e^{i\varphi_0}  \big( \varepsilon +(\gamma-1)\ell\big)^{1/2}
\big( \varepsilon +(\gamma+1)\ell\big)^{1/2}
=
\frac{e^{i2\varphi}}2\left(i p_\rho - \frac{p_\varphi}{\rho} +\rho\right) \left(- i p_\rho +\frac{p_\varphi}{\rho} +\rho\right)\,.
\end{equation}
Subsituting
$p_\rho=\pm \sqrt{2\varepsilon - \ell^2/\rho^2 -\rho^2 + 2\gamma \ell}$ from \eqref{veff} and $p_\varphi =\ell$, we
get the explicit solution
\begin{equation}\label{ho-trajectory}
\rho = \frac{|\ell|}{\sqrt{\varepsilon - 
(\varepsilon^2-\ell^2)^{1/2} \cos(2\varphi-\varphi_0)}} \, ,
\end{equation}
which, as expected, is the equation of an ellipse.

\medskip

\item
Trajectories of the Landau system: $p=1, q=0$.
Now, the equation of the trajectory becomes
\begin{equation}
e^{i\varphi_0}  
\big(\varepsilon +(\gamma+1)\ell\big)^{1/2}
=
e^{i\varphi} \frac1{\sqrt2}\left(- i p_\rho +\frac{p_\varphi}{\rho} +\rho\right)\,,
\end{equation}
and the solution is
\begin{equation}\label{lan-trajectory}
\rho = 
\frac{(2\varepsilon+4\ell)^{1/2}\cos(\varphi-\varphi_0) \pm
\sqrt{(2\varepsilon+4\ell)\cos^2(\varphi-\varphi_0) - 4 \ell}}{2} \, ,
\end{equation}
which is the polar equation of a circle with center at
$(\sqrt{\varepsilon/2+\ell} \cos \varphi_0,\sqrt{\varepsilon/2+\ell} \sin \varphi_0)$
and radius $\sqrt{\varepsilon/2}$. 

\end{itemize}
The trajectories for
other FD systems have more complicated expressions in polar coordinates, as can be seen in (\ref{sp2new}). However, the trajectories
can be expressed more easily in parametric form in Cartesian coordinates, as we will see in the next section.

%%%%%%%%%%%%%%%%%%%%
\subsection{Classical motion}

In order to find the motion of the classical FD system, we write the equations of motion for 
$\alpha^\pm,\,\beta^\pm$  as
\begin{equation}
\dot \alpha^\pm = \{\alpha^\pm,  h\} = \pm i(1+\gamma)\alpha^\pm,\qquad
\dot \beta^\pm = \{\beta^\pm,  h\} = \pm i(1-\gamma)\beta^\pm\,,
\end{equation}
which can be immediately integrated to give
\begin{equation}
\alpha^\pm(t) = e^{\pm i (1+\gamma) t}\alpha^\pm(0),\qquad
\beta^\pm(t) = e^{\pm i (1-\gamma) t}\beta^\pm(0)\,,
\label{abt}
\end{equation}
where the integration constants are $\alpha^\pm(0),\,\beta^\pm(0)\in \mathbb C$. 
The classical Hamiltonian given in (\ref{habbC}) is a constant of motion which, according to (\ref{abt}), takes the value 
\begin{equation} 
h = (1+\gamma)|\alpha^+(0)|^2 
+ (1-\gamma)|\beta^+(0)|^2 \, .
\end{equation}

The evolution
of $\alpha^\pm(t)$ and $\beta^\pm(t)$ given in (\ref{abt}) leads to the motion which, 
in principle can be expressed in terms of  
polar coordinates $\rho$ and $\varphi$, or equivalently, in terms of Cartesian coordinates ($x$, $y$). It happens that the formulas
of motion are much simpler in the Cartesian coordinates, so hereafter we will restrict to them.

\begin{figure}[htb]
\centering
\includegraphics[width=0.34\textwidth]{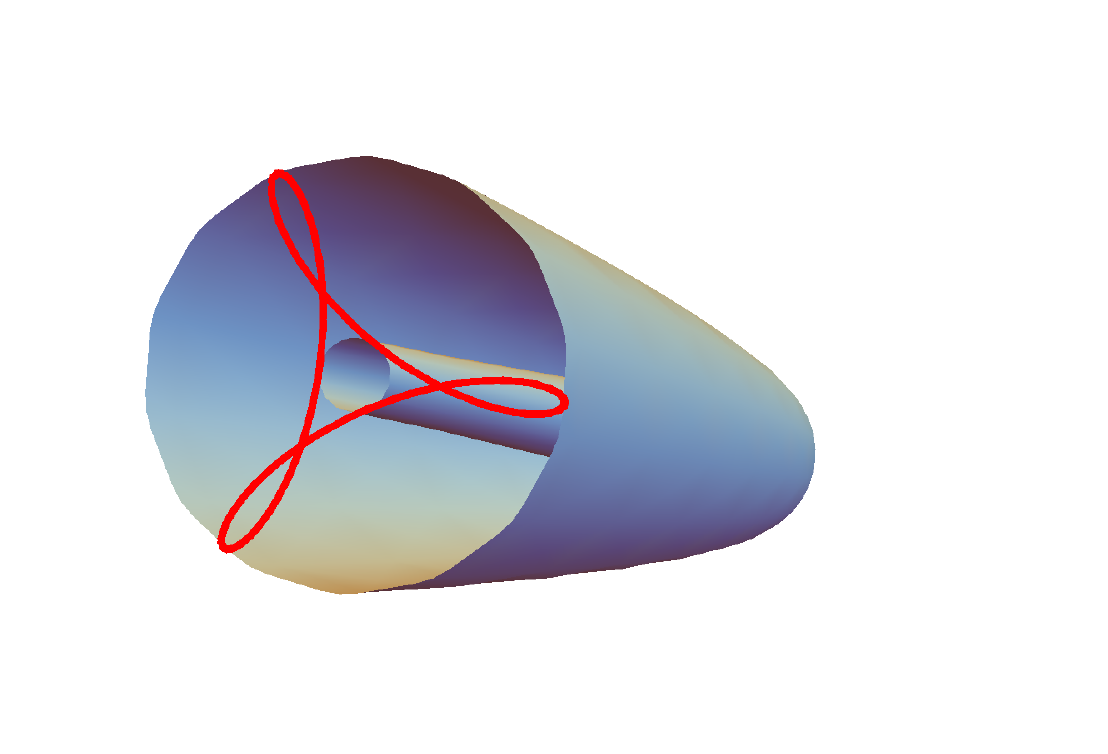}
\quad 
\includegraphics[width=0.27\textwidth]{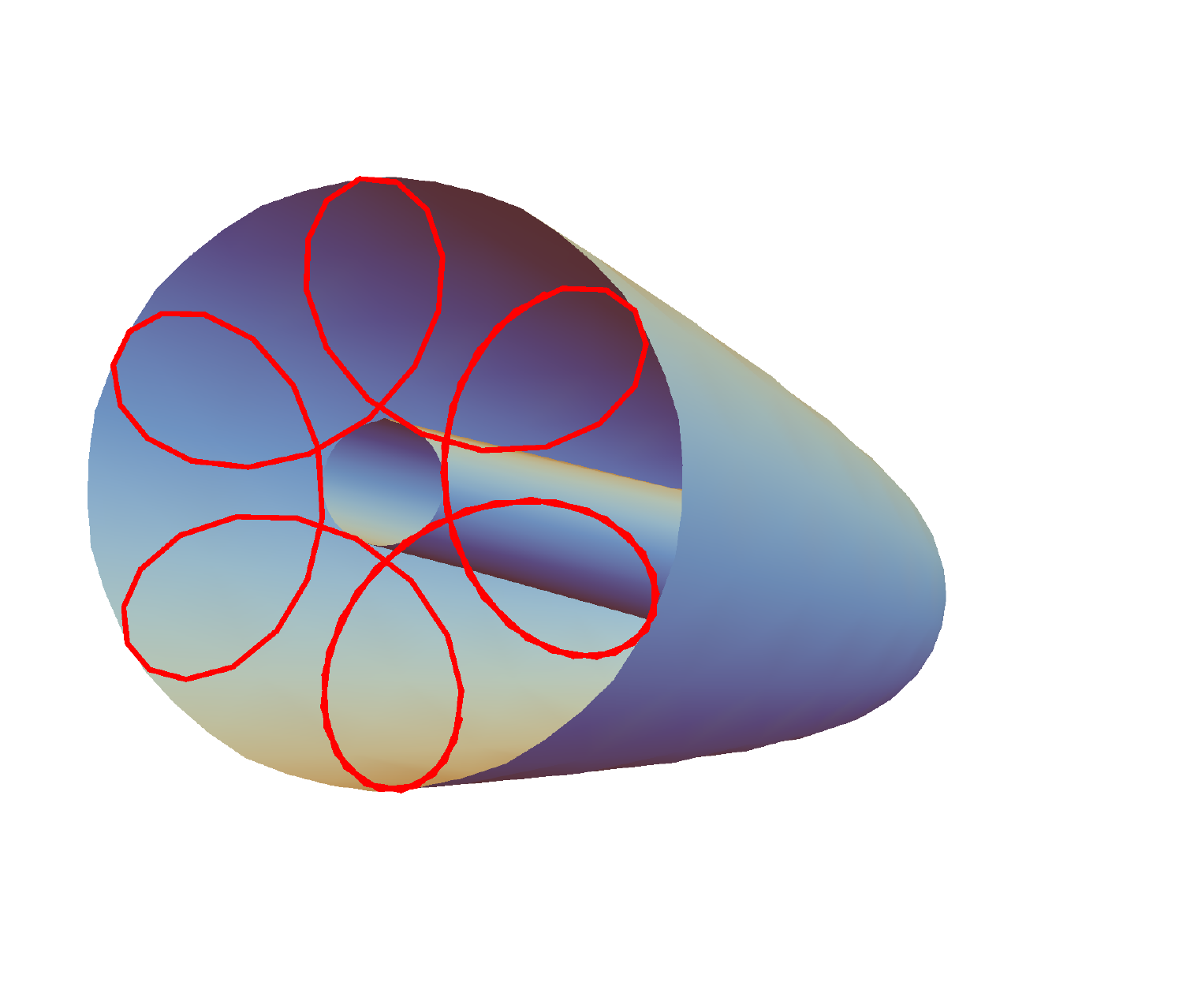} 
\qquad 
\includegraphics[width=0.28\textwidth]{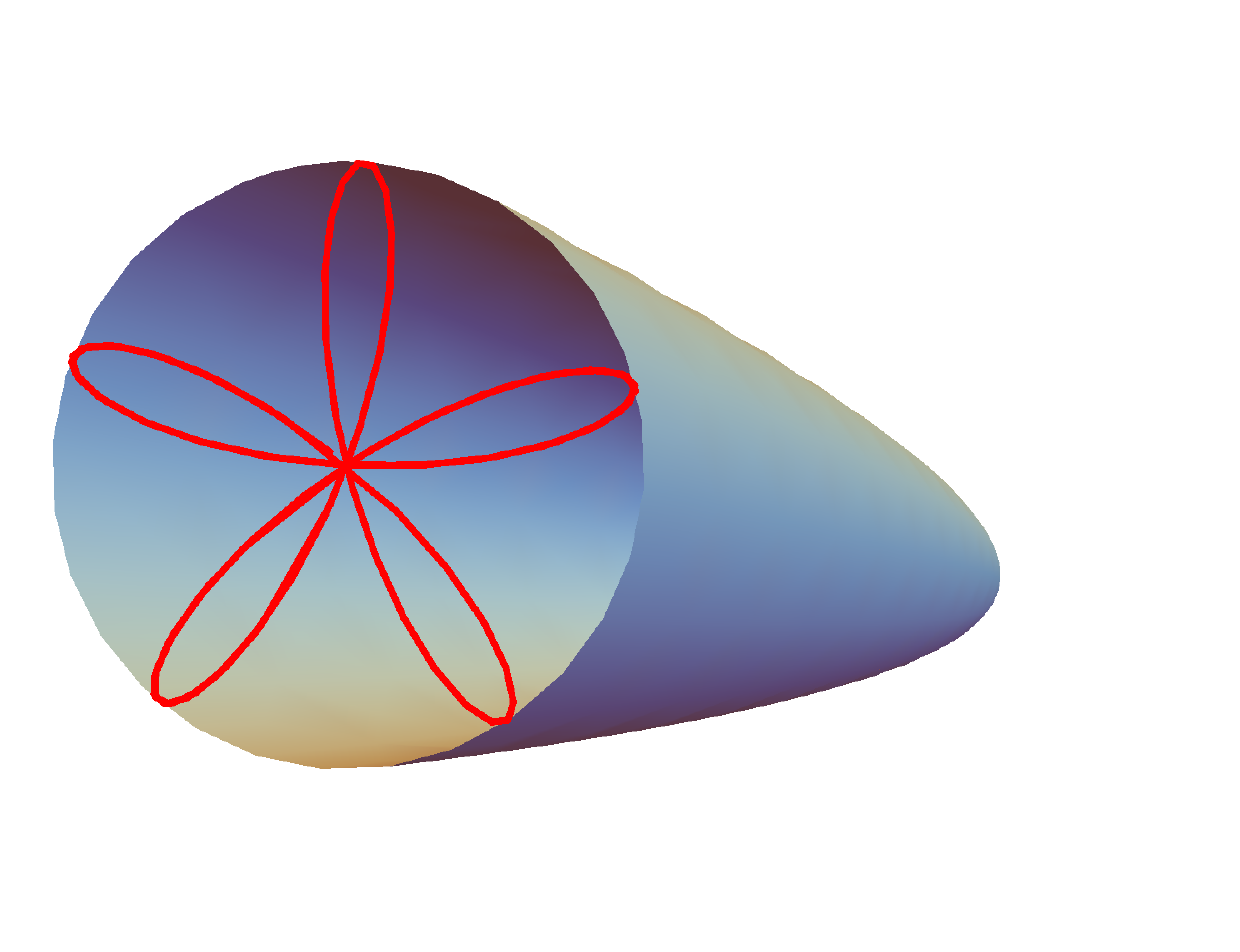}
\caption{\small Three plots of the classical trajectories \eqref{tracoor} of the classical FD system (in red) with the surfaces of the effective potentials. On the left $\{\gamma=1/3, p+q=3, \ell=96, \alpha_0=10, \beta_0=14,\theta_1= \theta_2=0 \}$, on the center 
$\{\gamma=2/3, p+q=6, \ell=125, \alpha_0=10, \beta_0=15, \theta_1=2, \theta_2=3\} $, and on the right $\{\gamma=1/5, p+q=5, \ell=0, \alpha_0=12, \beta_0=12, \theta_1=2, \theta_2=1 \}$.
\label{trajectory}}
\end{figure}

Let us now express the ladder functions 
$\alpha^\pm,\,\beta^\pm$ in terms of Cartesian canonical coordinates: $(x,\,p_x,\,y,\,p_y)$. To do this, we recall the expressions of the polar momenta 
\begin{equation}
p_{\rho}=\dfrac{x}{\sqrt{x^2+y^2}}\,p_x + \dfrac{y}{\sqrt{x^2+y^2}}\,p_y\,,
\qquad p_{\varphi}=-y\,p_x+x\,p_y\,.
\label{prp}
\end{equation}
Substituting (\ref{prp}) in (\ref{bpmbC}), writing $e^{\pm i\,\varphi}$  in terms of Cartesian coordinates and after straightforward computations, we get the simple linear expressions
\begin{equation}
\alpha^\pm = \frac12(-p_y+x) \mp i\,\frac12(p_x+y),\qquad
\beta^\pm = \frac12(p_y+x) \pm i\,\frac12(-p_x+y)\, .\label{abxy}
\end{equation}
We write the initial conditions in the form 
$\alpha^\pm(0)= \alpha_0\, e^{\pm\,i \,\theta_1}$
and $\beta^\pm(0)=\beta_0 \,e^{\pm\,i \,\theta_2}$, with $\alpha_0,\,\beta_0\in \mathbb R^+$, and $\theta_1, \theta_2\in \mathbb R$.
This is equivalent to take the initial conditions  
\begin{eqnarray*}
p_x(0)=-\alpha_0 \sin{\theta_1}- \beta_0 \sin{\theta_2},\, && x(0)=\alpha_0 \cos{\theta_1}+\beta_0\cos{\theta_2}, \\
p_y(0)=- \alpha_0 \cos{\theta_1} + \beta_0 \cos{\theta_2}, && y(0)=- \alpha_0\sin{\theta_1}+\beta_0 \sin{\theta_2}.
\end{eqnarray*}
Then, using (\ref{abt}) and (\ref{abxy}), we arrive to the explicit equations of motion in
Cartesian coordinates:
\begin{equation}
\begin{array}{l}
x(t) =\alpha_0 \cos[(1+\gamma)t +\theta_1]+ \beta_0 \cos[(1-\gamma)t+\theta_2]\,,
\\[1.ex]
y(t) =  -\alpha_0 \sin[(1+\gamma)t+\theta_1 ]+ \beta_0 \sin[(1-\gamma)t+\theta_2]\,,
\end{array}\label{tracoor}
\end{equation}
\begin{equation}
\begin{array}{l}
p_x(t) = -\alpha_0  \sin[(1+\gamma)t +\theta_1]- \beta_0 \sin[(1-\gamma)t+\theta_2]\,,
\\[1.ex]
p_y(t) = - \alpha_0 \cos[(1+\gamma)t +\theta_1]+\beta_0 \cos[(1-\gamma)t +\theta_2]\,.
\end{array}\label{tramom}
\end{equation}
 From here we can study the following special situations:
\begin{itemize}
\item
For $\gamma=0$, we have the motion in a two dimensional isotropic HO potential.
The trajectories correspond to ellipses or circles.

\item
For $\gamma=1$, we have the motion of the Landau system in a constant magnetic field.
The trajectories \eqref{tracoor} correspond to circles whose centers depend on the initial conditions. 

\item
For $\gamma$ a rational number with $|\gamma|<1$, we already know that $(1+\gamma)/(1-\gamma)= p/q$ is also rational, and we have the motion of the superintegrable FD system with closed trajectories.
Some examples of these trajectories, together with the corresponding effective potentials, are shown in 
Figure~\ref{trajectory}. When the values of $\alpha_0$ and $\beta_0$ are equal, the effective potential given by (\ref{veff}) has no singularity because the angular momentum $p_{\varphi}=|\beta^+(0)|^2-|\alpha^+(0)|^2$ is zero. 
This can be seen in the last graphic of 
Figure~\ref{trajectory}. When they have different values, the angular momentum is different from zero and the effective potential has a singularity. This can be appreciated in the first two graphics of Figure~\ref{trajectory}. 
The number of the lobes of the trajectories depends on the values of $p$ and $q$ and it is given by $p+q$, because the fundamental period of the trajectories is given by $T=2 \pi/(p+q)$ due to  (\ref{sp2new}). The number of total turning points for the coordinate
$\rho$ is $2(p+q)$.
\end{itemize}
%%%%%%%%%%%%%%%%%%%%%%%%%

%%%%%%%%%%%%%%%%%%
\subsection{Symmetries of the classical FD system}
%%%%%%%%%%%%%%%%%%

 It is well known that, in general, any constant of motion $G({\rm \bf{q}},{\rm\bf{ p}})$ produce a type of infinitesimal canonical transformations on the phase-space, which lead to transformations of the classical trajectories:  
 $G({\rm \bf{q}},{\rm\bf{ p}})$  can be considered as a generator of an infinitesimal canonical transformation, such that any  function $u({\rm \bf{q}},{\rm\bf{ p}})$  is changed as follows  
 \begin{equation}
 \frac{d\, u}{d\, \eta} = \{ u, G\},
 \label{uepsilon}
 \end{equation}
 where  $\eta$ is a continuous parameter. In principle, by integrating equation  (\ref{uepsilon}),
a finite canonical transformation,  $u(\eta)$, is obtained. A formal solution can be found by expanding $u(\eta)$ in a Taylor series about the initial conditions  \cite{goldstein}.
 
For the problem we are studying, it is convenient to start by computing the changes generated by  ${\cal S}^\pm$ and ${\cal L}$ given by \eqref{symFDclassical}-\eqref{symFDclassicalzz} on the functions $\alpha^\pm$ and $\beta^\pm$, since the canonical variables $(x, y, p_x, p_y)$ can be expressed in terms of $\alpha^\pm$ and $\beta^\pm$ according to (\ref{abxy}). We can evaluate the following: 

\begin{itemize}

 \item[$\star$]
 Infinitesimal action of ${\cal L}$:
 \begin{equation}
 \frac{d\, \alpha^\pm}{d\, \eta}= \{\alpha^\pm,{\cal L}\} =\mp i \alpha^\pm,\quad
  \frac{d\, \beta^\pm}{d\, \eta}= \{\beta^\pm,{\cal L}\} = \pm i \beta^\pm\, .
  \label{al}
 \end{equation}
 
  \item[$\star$]
 Infinitesimal action of ${\cal S}^+$:
  \begin{equation}
 \frac{d\, \alpha^+}{d\, \eta}=  i q (\alpha^-)^{q-1}(\beta^+)^p,\quad
 \frac{d\, \alpha^-}{d\, \eta}=0,\quad
 \frac{d\, \beta^+}{d\, \eta}=0,\quad
 \frac{d\, \beta^-}{d\, \eta}=  - i p (\alpha^-)^{q}(\beta^+)^{p-1}\, .
  \label{asp}
 \end{equation}

   \item[$\star$]
 Infinitesimal action of ${\cal S}^-$:
  \begin{equation}
 \frac{d\, \alpha^+}{d\, \eta}= 0,\quad
 \frac{d\, \alpha^-}{d\, \eta}= -i q (\alpha^+)^{q-1}(\beta^-)^p,\quad
 \frac{d\, \beta^+}{d\, \eta}= i p (\alpha^+)^{q}(\beta^-)^{p-1},\quad
 \frac{d\, \beta^-}{d\, \eta}= 0\, .
  \label{asm}
 \end{equation}
\end{itemize}

In the sequel we will deal with the three special cases we have already considered in the quantum context: HO, Landau and rational FD systems. We will show that the integration of
these differential equations leads to the finite action of symmetry transformations.
In the case of the HO and Landau systems such finite transformations are linear and
we will find then explicitly. However, in the generic rational FD system we will
only be able to find the explicit formulas for some special cases which are essentially nonlinear. 

\subsubsection{Harmonic oscillator ($q=p=1$ or  $\gamma=0$)}
We introduce the new (real) constants of motion 
\begin{equation} {\cal S}_1=({\cal S}^+ +{\cal S}^-)/2,\qquad  {\cal S}_2=({\cal S}^+- {\cal S}^-)/2i\,, \qquad {\cal S}={\cal L}/2,
\end{equation} 
in terms of 
$ \{{\cal S}^{\pm},{\cal L} \}$ given by \eqref{symFDclassical}-\eqref{symFDclassicalzz}, for $q=p=1$. These new constants of motion $\{{\cal S}_1, {\cal S}_2, {\cal S} \}$ close  the   Lie algebra $su(2)$ \cite{goldstein}, with Poisson brakets
\begin{equation}
\{{\cal S},{\cal S}_1\}= {\cal S}_2, \qquad \{{\cal S},{\cal S}_2\}= -{\cal S}_1, \qquad\{{\cal S}_1,{\cal S}_2\}=  {\cal S}\, .
\label{algebraho}
\end{equation} 

The angular momentum $2{\cal S}$ generates rotations of the classical trajectories, while ${\cal S}_1$ and  ${\cal S}_2$ give a type of transformation changing the shape of the trajectories. The finite transformations for these generators can be obtained by integrating the differential equations
 (\ref{al})-(\ref{asm}). The results are the following:
 \begin{itemize}
  \item[$\star$] Finite action of ${\cal S}_1$:
 \begin{equation}\label{lclas111}
 \begin{array}{ll}
 x'= x \cos \eta/2 + p_x \sin \eta/2,\quad &
 y'= y \cos \eta/2 - p_y \sin \eta/2,
 \\[2.ex]
 p'_x= p_x \cos \eta/2 - x \sin \eta/2,\quad &
 p'_y= p_y \cos \eta/2 + y \sin \eta/2\,. \end{array}
 \end{equation}
 
  \item[$\star$] Finite action of ${\cal S}_2$:
\begin{equation}\label{lclas222}
 \begin{array}{ll}
 x'= x \cos \eta/2 + p_y \sin \eta/2,\quad &
 y'= y \cos \eta/2 + p_x \sin \eta/2,
 \\[2.ex]
 p'_x= p_x \cos \eta/2 - y \sin \eta/2,\quad &
 p'_y= p_y \cos \eta/2 - x \sin \eta/2\,. \end{array}
 \end{equation}
 
   \item[$\star$] Finite action of ${\cal S}$:
\begin{equation}\label{lclas}
 \begin{array}{ll}
 x'= x \cos \eta/2 - y \sin \eta/2,\quad &
 y'= y \cos \eta/2 +  x \sin \eta/2,
 \\[2.ex]
 p'_x= p_x \cos \eta/2 - p_y \sin \eta/2,\quad &
 p'_y= p_y \cos \eta/2 + p_x \sin \eta/2\,. \end{array}
 \end{equation}

 \end{itemize}
 
\begin{figure}[htb]
\centering
\includegraphics[width=0.30\textwidth]{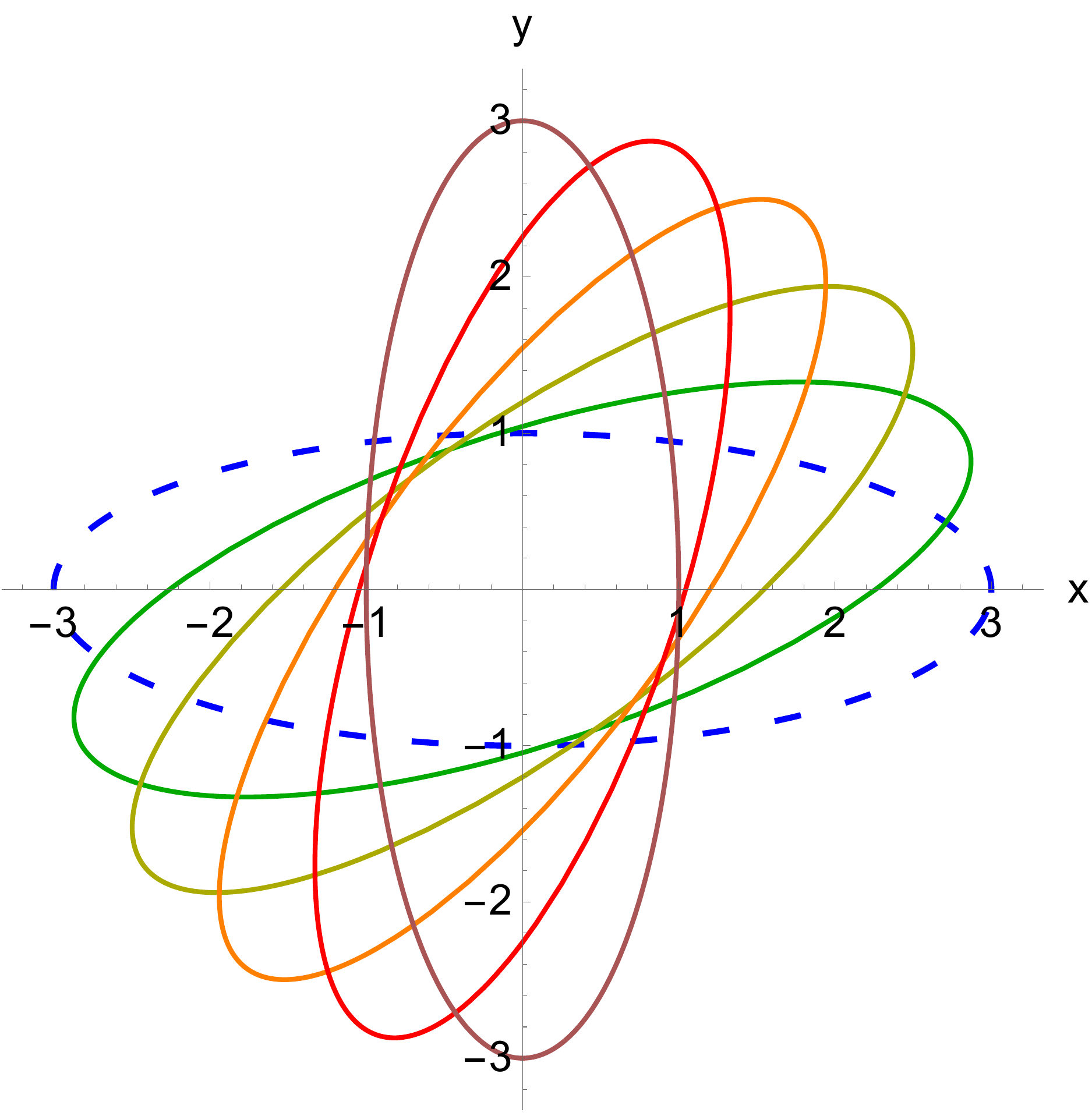}
\quad 
{\includegraphics[width=0.30\textwidth]{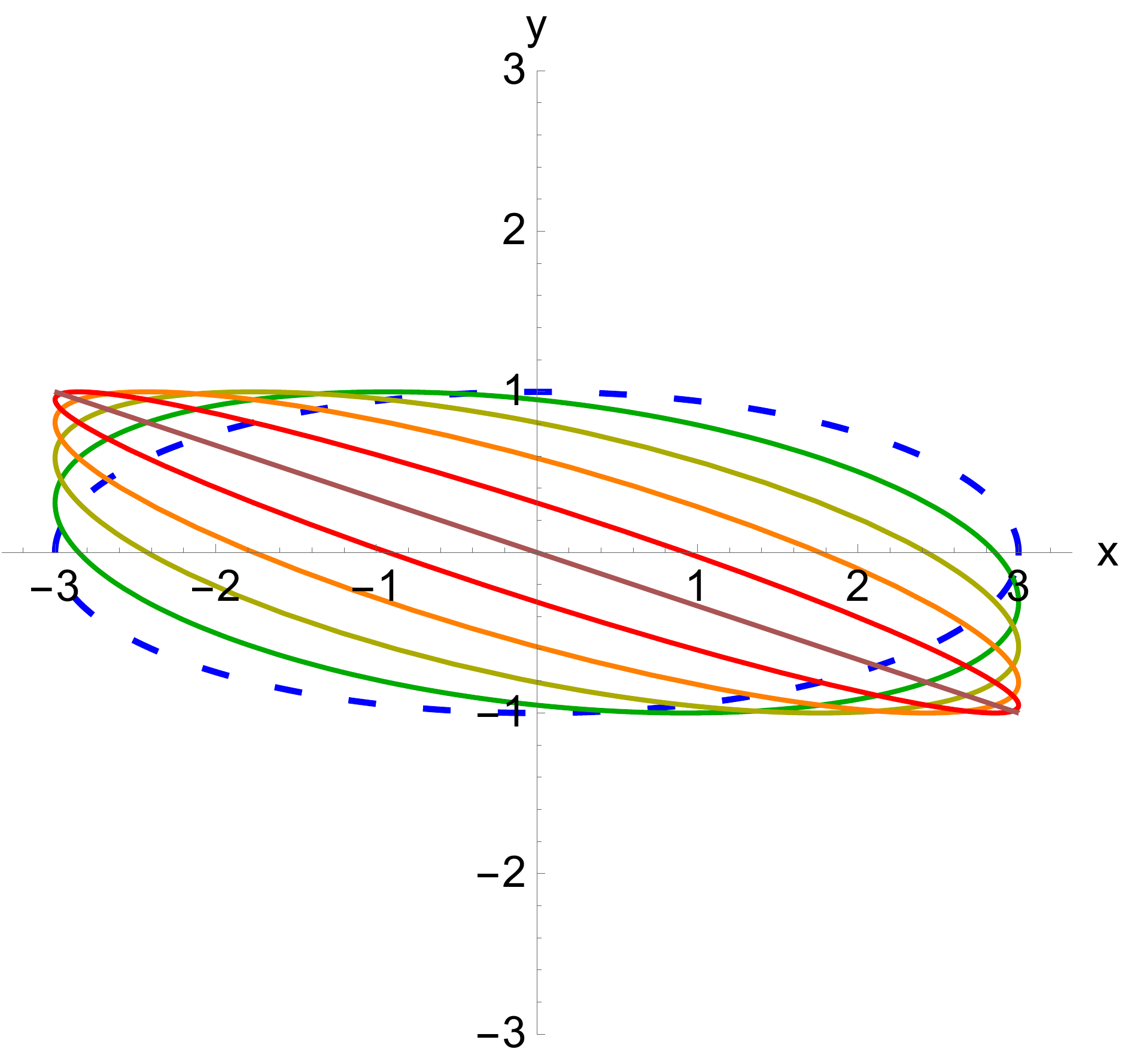}}
\quad 
\includegraphics[width=0.30\textwidth]{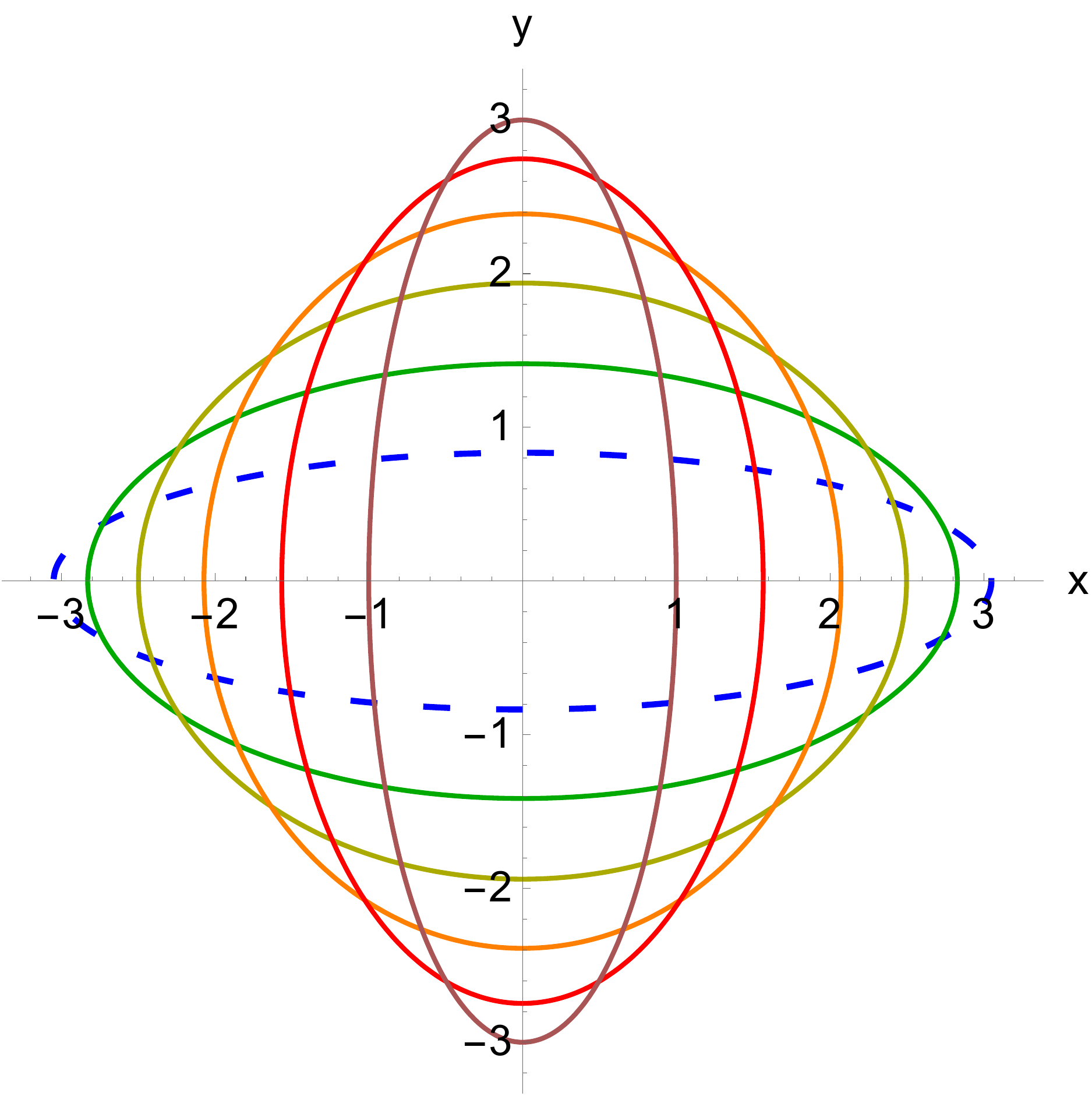}
\caption{\small   Plot of the action of the symmetry group elements on the classical trajectories of the harmonic oscillator: on the left they are generated by ${\cal S} $ \eqref{lclas}, in the center by ${\cal S}_1$ \eqref{lclas111}, and on the right by ${\cal S}_2$ \eqref{lclas222}; the relevant parameters are chosen to be $\{\gamma=0, \alpha_0=1, \beta_0=2, \theta_1=0, \theta_2=0 \}$.
\label{trajectoryLS}}
\end{figure}

The effects of all these transformations (classical symmetries) for different values of $\eta$ on the trajectories can be seen in Figure~\ref{trajectoryLS}. In these plots, the dashed lines correspond to the initial trajectory
($\eta=0$).

The transformations generated by ${\cal S}_1$ and  ${\cal S}_2$ leave the Hamiltonian invariant but they change the value of the angular momentum.  This means that, under these transformations, the effective potential changes, but the energy is conserved, and therefore they  may be considered as classical analogs of the quantum mechanical shift operators.

%%%%%%%%%%%%%%%%%%%%%%%%

%%%%%%%%%%%%%%%%%%
\subsubsection{Landau system ($q=1, p=0$ or $\gamma=1$)}

For this case, the constant of motions given by (\ref{symFDclassical}) take the form
\begin{equation}
{\cal M}= \alpha^+\alpha^-\,,\qquad {\cal N}=  \beta^+\beta^-\,, 
\qquad {\cal S}^- = \beta^-,\qquad {\cal S}^+ =\beta^+\,.
\label{symlandauclas}
\end{equation}
 We introduce again real constants of motion in terms of ${\cal S}^{\pm}$ given by (\ref{symlandauclas}) and ${\cal L}$: 
\begin{equation}
{\cal S}_1=({\cal S}^+ +{\cal S}^-)/2\,,\qquad  
{\cal S}_2=({\cal S}^+- {\cal S}^-)/2i\,,\qquad{\cal S}={\cal L}={\cal N}-{\cal M}. 
\label{symlandauclassss}
\end{equation}
They satisfy 
\begin{equation}
\{{\cal S},{\cal S}_1\}={\cal S}_2, \qquad \{{\cal S},{\cal S}_2\}= -{\cal S}_1, \qquad\{{\cal S}_1,{\cal S}_2\}= -\frac 12 .
\label{algebralandau}
\end{equation} 

These constants of motion are generators of  symmetries which leave invariant the Hamiltonian.  The finite transformations for these generators are obtained by integrating the differential equations (\ref{al})-(\ref{asm}):
\begin{itemize}
 \item[$\star$] Finite  action of ${\cal S}_1$:
\begin{equation}\label{harto1}
x'=x,\qquad y'=y+\frac{\eta}{2},\qquad p'_x=p_x-\frac{\eta}{2},\qquad p'_y=p_y\,.
\end{equation}
 \item[$\star$] Finite action of ${\cal S}_2$:
\begin{equation}\label{harto2}
x'=x-\frac{\eta}{2},\qquad y'=y,\qquad p'_x=p_x,\qquad p'_y=p_y-\frac{\eta}{2}\,.
\end{equation}
\end{itemize}
The action of ${\cal S}$ has the same form as in (\ref{lclas}).

\begin{figure}[htb]
\centering
\includegraphics[width=0.30\textwidth]{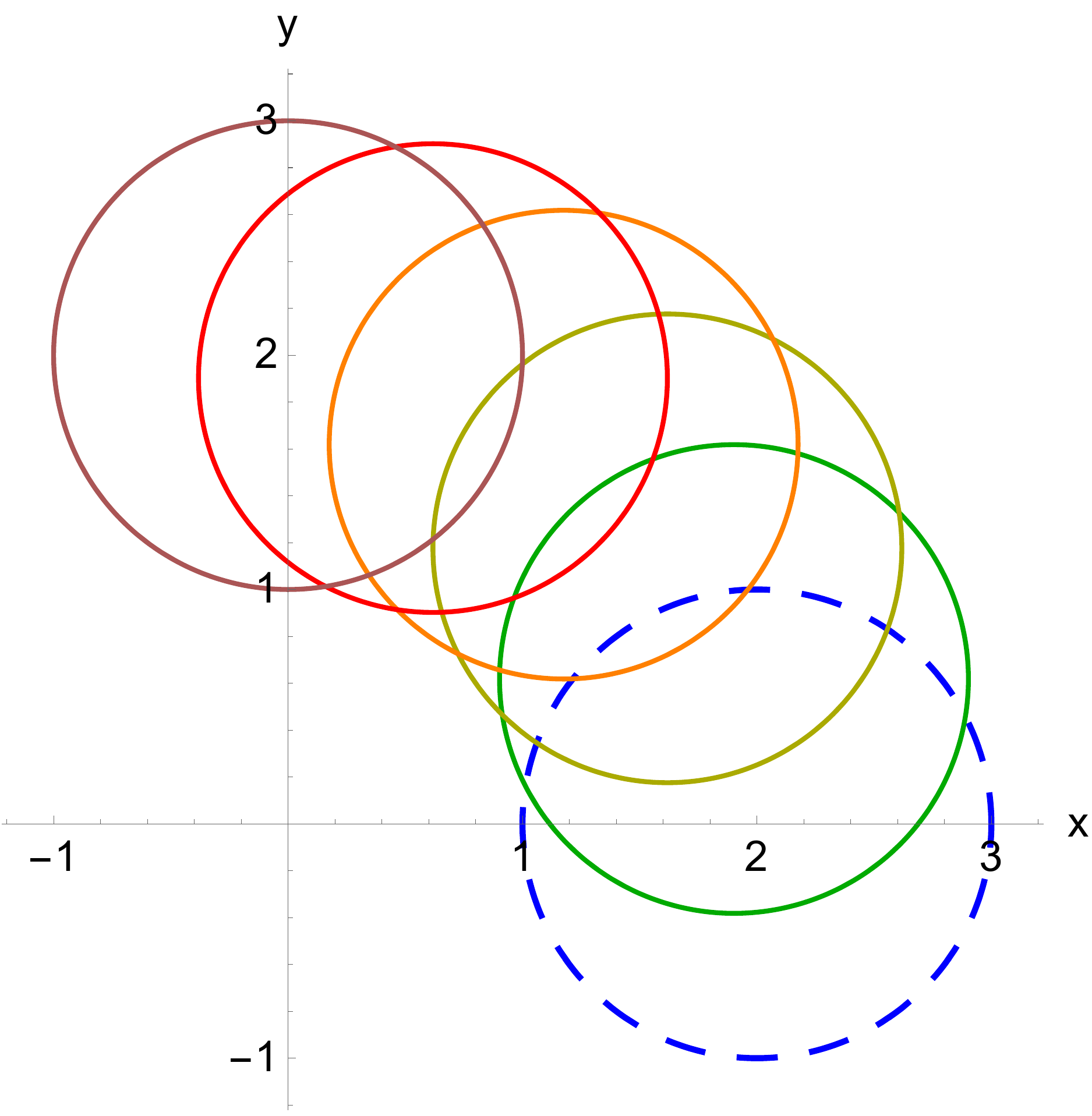}
\quad 
\includegraphics[width=0.25\textwidth]{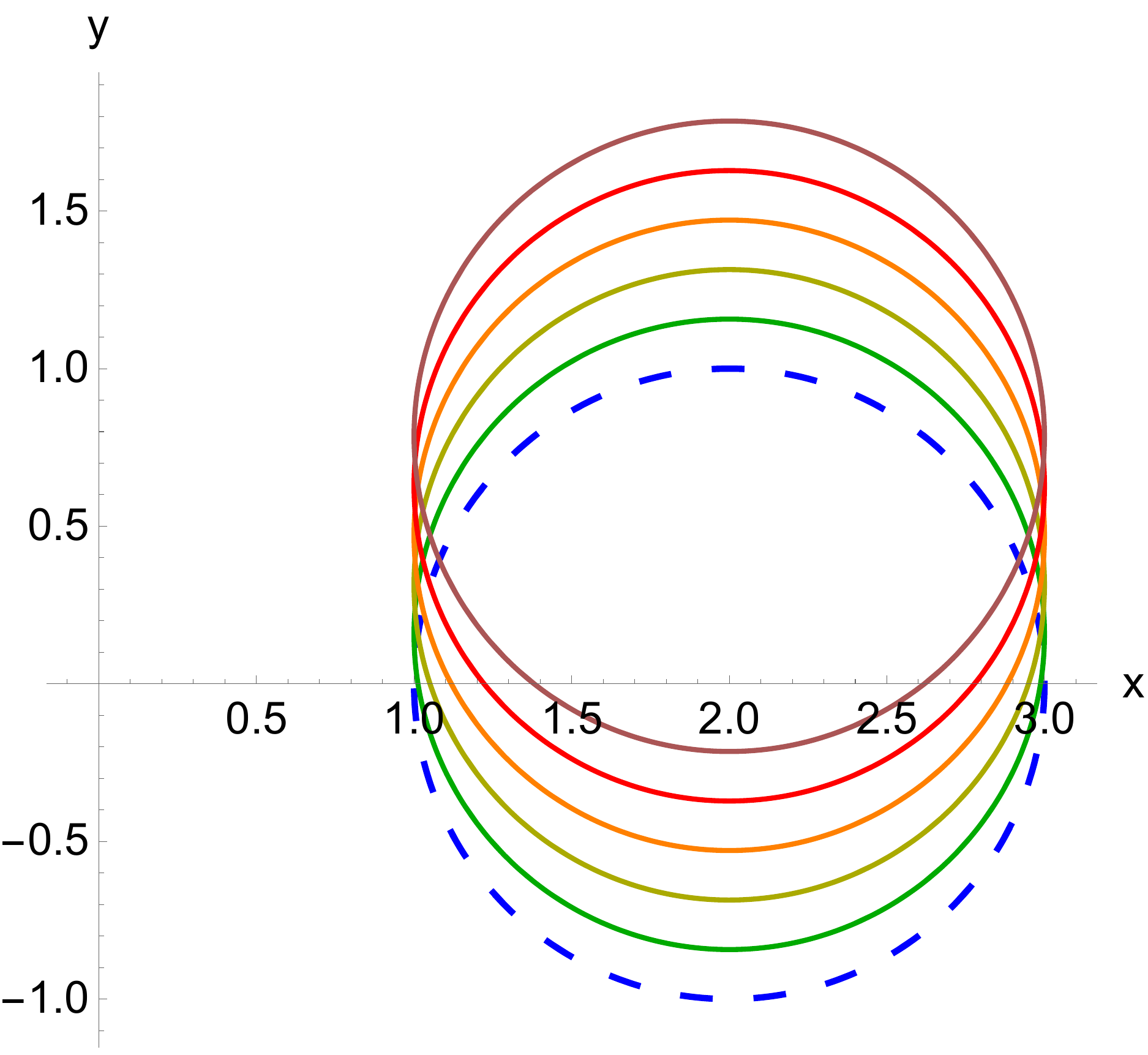}
\quad 
{\includegraphics[width=0.25\textwidth]{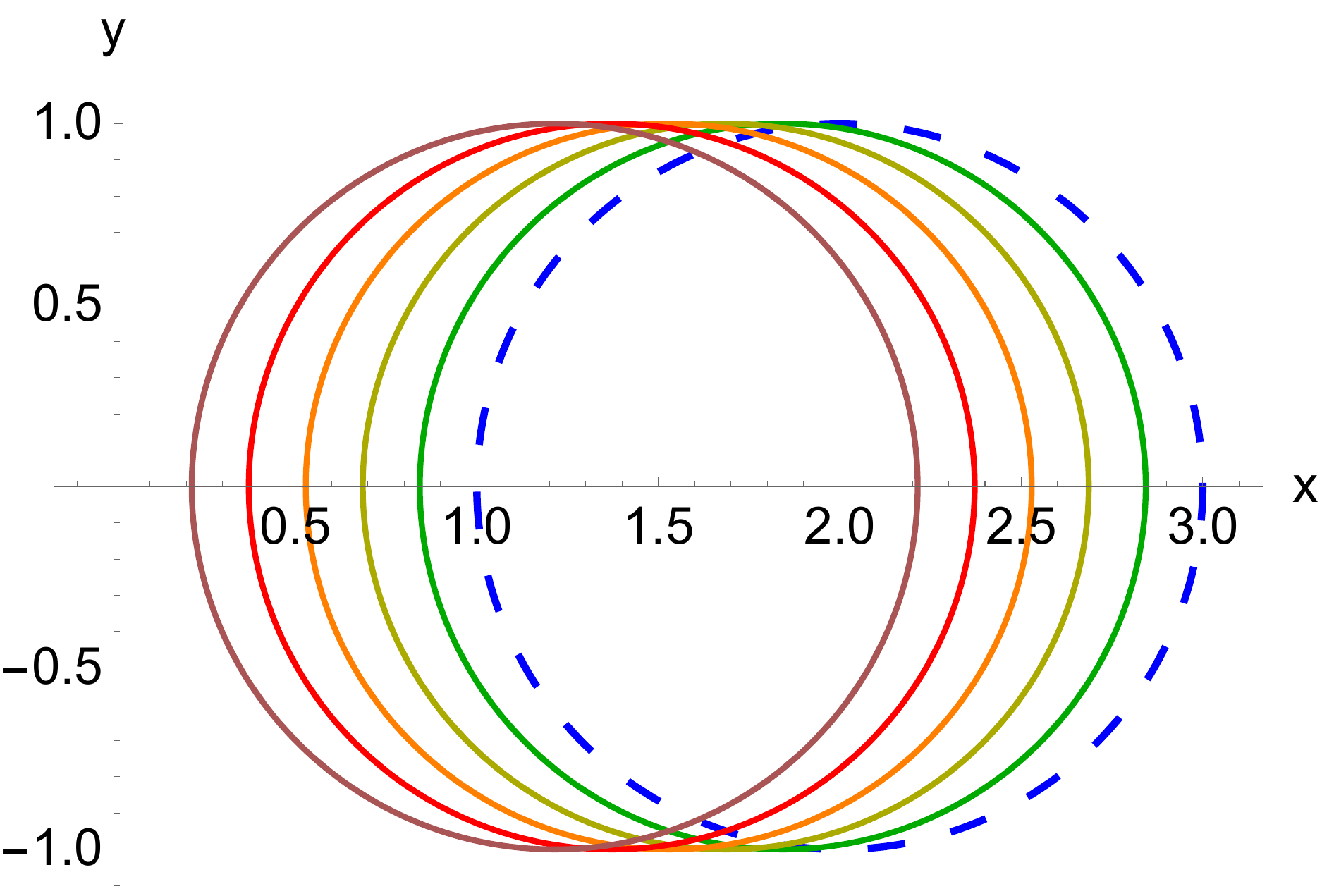}}
\caption{\small Plot of the action of the symmetry group elements on the classical trajectories of the Landau system: on the left they are generated by  ${\cal S} $ \eqref{lclas}, in the center by  ${\cal S}_1 $ \eqref{harto1}, and on the right by ${\cal S}_2$ \eqref{harto2};  the relevant parameters are chosen to be  $\{\gamma=1, \alpha_0=1, \beta_0=2, \theta_1=0, \theta_2=0 \}$.
\label{trajectorySLandau}}
\end{figure}

The effect of all these symmetry transformations for different values of $\eta$ on the trajectories can be seen in Figure~\ref{trajectorySLandau}. 
The value $\eta=0$ corresponds to the initial motion, and it is shown in Figure~\ref{trajectorySLandau} by dashed line.

%%%%%%%%%%%%%%%%%%%%%%%%

%%%%%%%%%%%%%%%%
\subsubsection{Rational FD system for arbitrary $q, p$}

For the generic rational FD system we consider  the constants of motion already introduced in \eqref{symlandauclassss}.
Again, ${\cal S}$ is the
angular momentum, and therefore it generates rotations, as described in the HO and Landau
subsection. Thus, we will concentrate on the action of ${\cal S}_1$ and ${\cal S}_2$.

\begin{figure}[htb]
\centering
\includegraphics[width=0.45\textwidth]{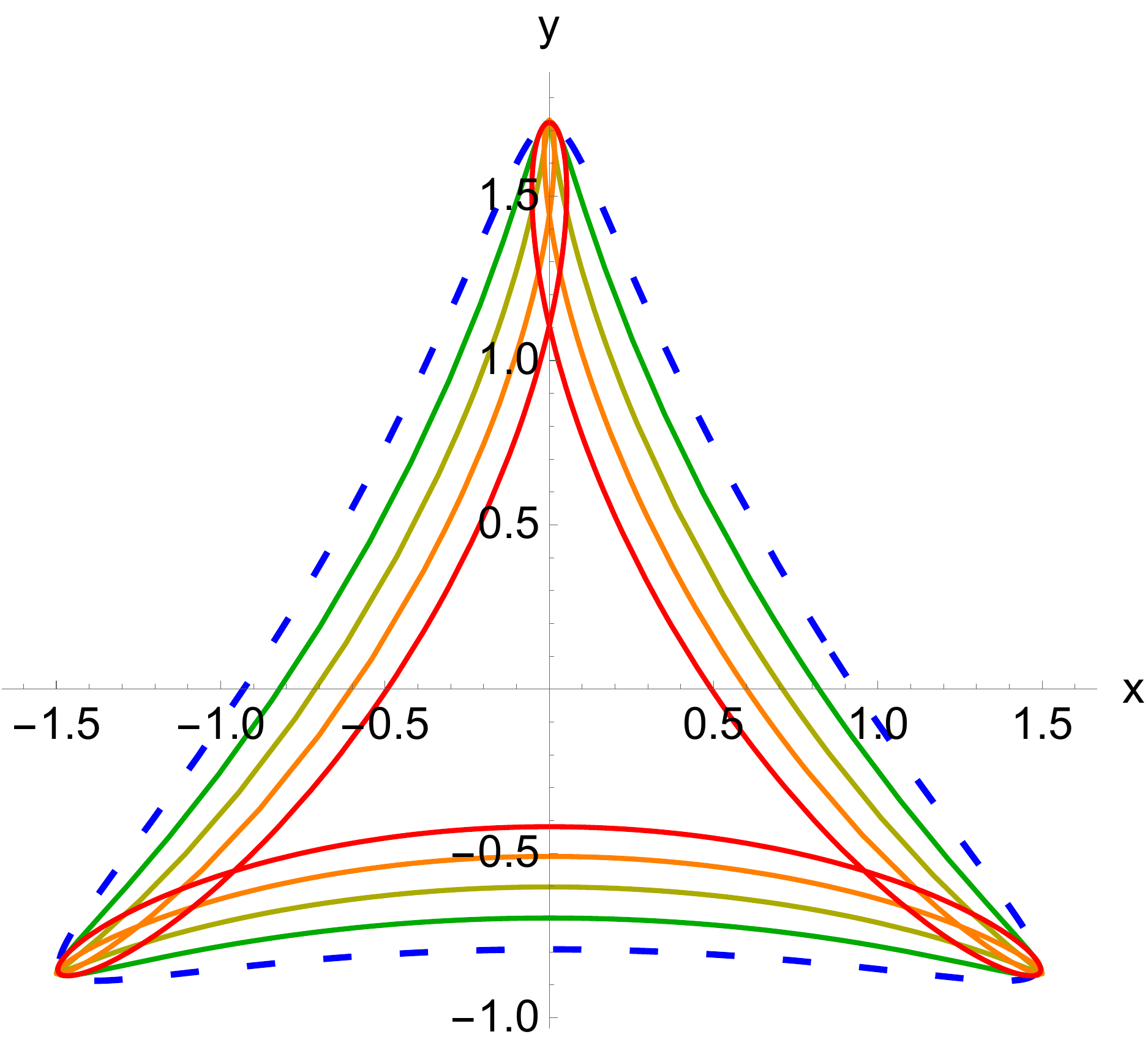}
\qquad \qquad
\includegraphics[width=0.4\textwidth]{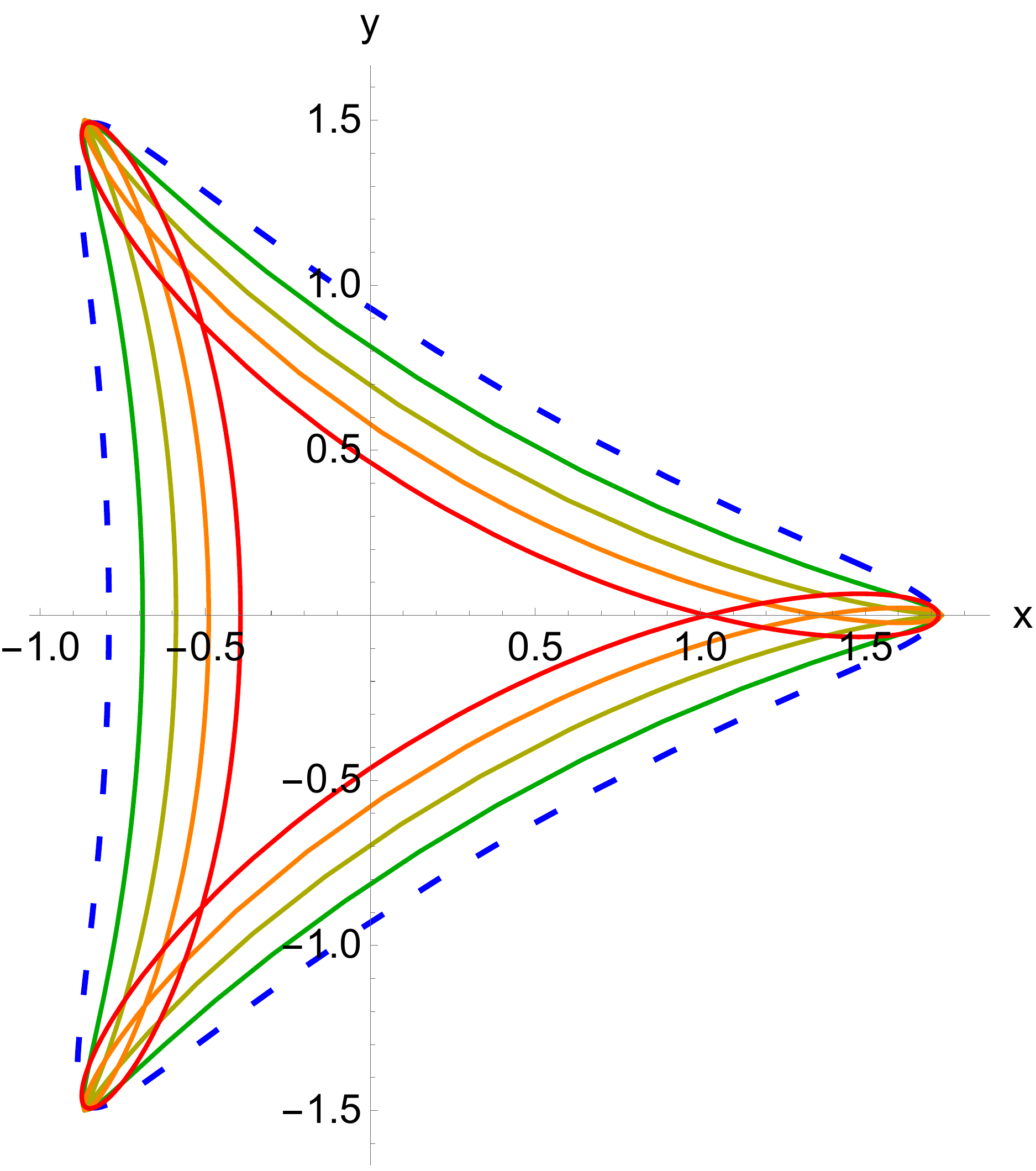}
\caption{\small   Different motions connected by the symmetries generated by ${\cal S}_1$
(left) and by ${\cal S}_2$ (right) corresponding to the energy $\varepsilon=2$ and $\gamma=1/3, q=1, p=2, c_2=0.5, \theta_2=\pi/8$. The initial motion is represented
by the dotted blue curve.
\label{Energia_constant}}
\end{figure}

 \begin{itemize}
 \item[\bf (A)]  Finite action of ${\cal S}_1$: we have to solve the following nonlinear equations:
  \begin{equation}
  \begin{array}{l}
 \dfrac{d\, \alpha^+}{d\, \eta}=  i \dfrac{q}{2} (\alpha^-)^{q-1}(\beta^+)^p,\qquad
 \dfrac{d\, \alpha^-}{d\, \eta}=-i \dfrac{q}{2} (\alpha^+)^{q-1}(\beta^-)^p,
 \\[2.Ex]
 \dfrac{d\, \beta^+}{d\, \eta}=i \dfrac{p}{2} (\alpha^+)^{q}(\beta^-)^{p-1},\qquad
 \dfrac{d\, \beta^-}{d\, \eta}= - i \dfrac{p}{2} (\alpha^-)^{q}(\beta^+)^{p-1}\, .
  \label{aspfd}
  \end{array}
 \end{equation}
 It is quite difficult to find the general solution of $\alpha^{\pm}$ and $\beta^{\pm}$ for any value of $q$ and $p$, but it is possible  to get some special solutions. 
Let us propose the following polar-type ansatz for the solutions of (\ref{aspfd})
 \begin{equation}
 \alpha^{\pm}(\eta)=\rho_1(\eta) \, e^{\pm i\theta_1(\eta)}\,,\qquad \beta^{\pm}(\eta)=\rho_2(\eta) \, e^{\pm i\theta_2(\eta)}\,,
 \label{proab}
 \end{equation}
where $\rho_1,\rho_2$ and $\theta_1,\theta_2$ are real functions depending on the group
parameter $\eta$.
 Substituting in (\ref{aspfd}), we arrive to the following equations
\begin{equation}
  \begin{array}{ll}
 \dfrac{d\, \rho_2}{d\, \eta}=  - \dfrac{p}{2}\, \rho_1^q  \,\rho_2^{p-1}\,\sin(q\theta_1-p\theta_2),& \qquad 
 \dfrac{d\,\theta_2}{d\, \eta}=  \, \,\dfrac{p}{2}\, \rho_1^q  \,\rho_2^{p-2}\,\cos(q\theta_1-p\theta_2),
 \\[2.Ex]
\dfrac{d\, \rho_1}{d\, \eta}=  \,\,\dfrac{q}{2}\, \rho_1^{q-1}  \,\rho_2^{p}\,\sin(q\theta_1-p\theta_2),& \qquad 
 \dfrac{d\,\theta_1}{d\, \eta}= \, \, \dfrac{q}{2}\, \rho_1^{q-2}  \,\rho_2^{p}\,\cos(q\theta_1-p\theta_2)\, .
  \label{eqnonab}
  \end{array}
 \end{equation}
 
They lead to  energy conservation for the classical FD system: $p \rho_1^2 + q\rho_2^2=c_1=\varepsilon$. Now, let us consider two special cases: 
\medskip

{\bf (A1)} If $q\theta_1-p\theta_2=0$,
then from \eqref{eqnonab} it follows that $\rho_1$ and $\rho_2$ are constants satisfying ${\rho_1}/{\rho_2}={q}/{p}$  and $\theta_1$ and $\theta_2$ are
linear functions of $\eta$ given by
\begin{equation}
\theta_1(\eta) =\frac{q}2 \rho_1^{q-2} \rho_2^{p}\,\eta +\phi_1,\qquad
\theta_2(\eta) =\frac{p}2 \rho_1^{q} \rho_2^{p-2}\, \eta +\phi_2\,,\qquad
\end{equation}
where the constants $\phi_1,\phi_2$ also satisfy $\phi_1/\phi_2 = p/q$.
In summary, we have integrated the action of the symmetry on the points characterized
by $\alpha^\pm= \rho_1\, e^{\pm i \phi_1},\  \beta^\pm =  \rho_2\, e^{\pm i \phi_2}$,
such that $\rho_1/\rho_2 = q/p$ and $\phi_1/\phi_2 = p/q$. It can be shown that this kind of transformations acting on these points give  the same trajectory as their corresponding motion.
\medskip

{\bf (A2)}
 If $q\theta_1-p\theta_2=\pi/2$, then $\theta_1$ and $\theta_2$ are constants and $\rho_1$ and $\rho_2$ are the functions of $\eta$. For example, we can obtain the explicit expressions if $q=1$ and $p=2$:
\begin{equation}
 \rho_1(\eta)=\sqrt{\frac{c_1}{2}} \,{\rm tanh}\left[{\sqrt{\frac{c_1}{2}}\,\eta+c_2}\right],\qquad \rho_2(\eta)=\sqrt{{c_1}} \,{\rm sech}\left[{\sqrt{\frac{c_1}{2}}\,\eta+c_2}\right],
 \label{rho12}
 \end{equation}
where $c_1$ and $c_2$ are integration constants. Then, we can express $\alpha^{\pm}$ and $\beta^{\pm}$ in terms of the transformation parameter $\eta$ as:
\begin{equation}
 \alpha^{\pm}(\eta)=\sqrt{\frac{c_1}{2}} \,{\rm tanh} \left[{\sqrt{\frac{c_1}{2}}\,\eta+c_2}\right] \, e^{\pm i\theta_1},
 \quad 
 \beta^{\pm}(\eta)=\sqrt{{c_1}} \,{\rm sech} \left[{\sqrt{\frac{c_1}{2}}\,\eta+c_2}\right]\, e^{\pm i\theta_2}.
 \label{ab12}
 \end{equation}

Finally, we express the finite action of $S_1$ as:
\begin{equation}
  \begin{array}{l}
x'=\sqrt{ {c_1} }\,{\rm sech}\left[ {\sqrt{\dfrac{c_1}{2}}\,\eta+c_2} \right]\,\cos \theta_2+\sqrt{\dfrac{c_1}{2}} \,{\rm tanh}\left[ {\sqrt{\dfrac{c_1}{2}}\,\eta+c_2} \right]\, 
\cos \left(2\theta_2+\dfrac{\pi}{2} \right)\,,
 \\[2.Ex]
y'=\sqrt{ {c_1} }\,{\rm sech}\left[ {\sqrt{\dfrac{c_1}{2}}\,\eta+c_2} \right]\,\sin \theta_2-\sqrt{\dfrac{c_1}{2}} \,{\rm tanh}\left[ {\sqrt{\dfrac{c_1}{2}}\,\eta+c_2} \right]\,
\sin \left(2\theta_2+\dfrac{\pi}{2} \right)\,,
\\[2.Ex]
p'_x=-\sqrt{ {c_1} }\,{\rm sech}
\left[ {\sqrt{\dfrac{c_1}{2}}\,\eta+c_2} \right]\,\sin \theta_2-\sqrt{\dfrac{c_1}{2}} \,{\rm tanh} \left[ {\sqrt{\dfrac{c_1}{2}}\,\eta+c_2} \right]\, 
\sin \left(2\theta_2+\dfrac{\pi}{2} \right)\,,
\\[2.Ex]
p'_y=\sqrt{ {c_1} }\,{\rm sech}\left[ {\sqrt{\dfrac{c_1}{2}}\,\eta+c_2} \right] \,\cos \theta_2-\sqrt{\dfrac{c_1}{2}} \,{\rm tanh}\left[ {\sqrt{\dfrac{c_1}{2}}\,\eta+c_2} \right]\,
\cos \left(2\theta_2+\dfrac{\pi}{2} \right)\,.
  \label{finiteaction12}
  \end{array}
 \end{equation}
In Figure~\ref{Energia_constant} (left), we represent some examples of motions which are related by means of this finite action of the symmetry $S_1$. The initial points $\alpha^\pm(0)$ and $\beta^\pm(0)$ are fixed by \eqref{rho12} with $\eta=0$, and $q\theta_1-p\theta_2=\pi/2$.
Case (A2) is more interesting than (A1) because these symmetry transformations connect different motions.
\medskip

 \item[\bf (B)]
 Finite action of ${\cal S}_2=({\cal S}^+-{\cal S}^-)/2i$. The differential equations to be solved are:
  \begin{equation}
  \begin{array}{ll}
 \dfrac{d\, \alpha^+}{d\, \eta}= \dfrac{q}{2} (\alpha^-)^{q-1}(\beta^+)^p,&\qquad
 \dfrac{d\, \alpha^-}{d\, \eta}= \dfrac{q}{2} (\alpha^+)^{q-1}(\beta^-)^p,
  \\[2.Ex]
 \dfrac{d\, \beta^+}{d\, \eta}= - \dfrac{p}{2} (\alpha^+)^{q}(\beta^-)^{p-1},&\qquad
 \dfrac{d\, \beta^-}{d\, \eta}= -  \dfrac{p}{2} (\alpha^-)^{q}(\beta^+)^{p-1}\, ,
   \end{array}
  \label{asmfd}
 \end{equation}
which have the same difficulties as the symmetry ${\cal S}_1$.
Nevertheless, we can find particular solutions corresponding to the two cases 
 (A1) and (A2) considered above for ${\cal S}_1$. They are identical, except for
a rotation of $\pi/2$. Some examples of these transformations are shown in 
Figure~\ref{Energia_constant} (right).
\end{itemize}

%%%%%%%%%%%%%%%%%%%%%%%%

\sect{Coherent states}

It was shown in Section~2.2 that the quantum FD system has two independent sets
of ladder operators $a^\pm$ and $b^\pm$, which generate all the eigenfunctions from
the ground state. Therefore, it is quite natural to define the coherent states for this system as the eigenstates of both annihilation operators $a^-$ and $b^-$:
\begin{equation}\label{annop}
a^- |\alpha, \beta\, \rangle= \alpha \, |\alpha, \beta \, \rangle \,,\qquad
b^- |\alpha, \beta \, \rangle= \beta \,  |\alpha, \beta\, \rangle  \,,
\end{equation}
where $\alpha,\beta\in{\mathbb C}$.
As both type of operators commute,  we can write $|\alpha, \beta\rangle =|\alpha\rangle \otimes |\beta \rangle$, being $|\alpha\rangle$ and $|\beta\rangle$  coherent states of the usual harmonic oscillator. Therefore, we can write
$|\alpha, \beta \rangle$ as
\begin{equation}
|\alpha, \beta \, \rangle= \left(e^{-|\alpha|^2/2} \sum_{m=0}^{\infty} {\frac{\alpha^m}{\sqrt{m!}}|m\rangle}\right) \otimes 
\left(e^{-|\beta|^2/2} \sum_{n=0}^{\infty} {\frac{\beta^n}{\sqrt{n!}}|n\rangle}\right)\,.
\label{alphabeta}
\end{equation}
Now, we are interested in an explicit form for the coherent state wavefunction  analytically by substituting the differential realizations (\ref{abpma})-(\ref{abpmb}) of the annihilation operators $a^-, b^-$ in ({\ref{annop}}):
\begin{equation}
\Psi_{\alpha\,\beta}(\rho,\varphi)= K(\alpha,\beta) \,\rho^{1/2} \,e^{-\rho^2/2}\, e^{\rho\,(\alpha\,e^{-i\,\varphi}+\beta\,e^{i\,\varphi})}\,,
\label{psicoherent}
\end{equation}
where $K(\alpha,\beta)$ is a normalization constant that must be determined because
it will play an essential role later. The probability density is
\begin{equation}
|\Psi_{\alpha\,\beta}(\rho,\varphi)|^2= |K(\alpha,\beta)|^2 \,\rho \,e^{-\rho^2}\, e^{2\,\rho\,u(\alpha\,,\beta)\,\cos{(\varphi-\varphi_0)}}\,,
\label{quadpsicoherent}
\end{equation}
where $u(\alpha\,,\beta)=|\alpha+\beta^*|$. 
After imposing normalization of the coherent state
$$
\int_{\mathbb{R}^2} |\Psi_{\alpha\,\beta}(\rho,\varphi)|^2\ d \rho\, d\varphi=1,
$$ 
$|K(\alpha,\beta)|^2$ can be expressed in terms of $u(\alpha\,,\beta)$ and  the modified Bessel functions $I_1, I_0$ as
\begin{equation}
|K(\alpha,\beta)|^2=\frac{2}{\pi^{3/2}\,e^{u^2/2}\,\left(u^2\,I_1(u^2/2)+(u^2+1)I_0(u^2/2)\right)}\,.
\label{coefK}
\end{equation}

The time evolution of the eigenstates $|m\rangle\otimes|n\rangle$ of the FD Hamiltonian is 
\begin{equation}
e^{-it H}|m\rangle\otimes|n\rangle=e^{-it\varepsilon_{mn}}|m\rangle\otimes|n\rangle\,,
\end{equation}
and taking into account the eigenvalue equation for the FD system 
$H\,\Psi_{m,n}=\varepsilon_{m,n}\,\Psi_{m,n}$, with $\varepsilon_{m,n}=m(1+\gamma)+n(1-\gamma)+1$, we can write the time evolution of $|\alpha, \beta \rangle$:
\begin{eqnarray}\nonumber
|\alpha, \beta, t \rangle&=& e^{-i\,t}\left(e^{-|\alpha|^2/2} \sum_{m=0}^{\infty} {\frac{(\alpha\,e^{-i\,(1+\gamma)t})^m}{\sqrt{m!}}|m\rangle}\right) \otimes 
\left(e^{-|\beta|^2/2} \sum_{n=0}^{\infty} {\frac{(\beta\,e^{-i\,(1-\gamma)t})^n}{\sqrt{n!}}|n\rangle}\right)\,
\\[2.ex]
&=&e^{-i\,t}|\alpha\,e^{-i\,(1+\gamma)t}, \beta\,e^{-i\,(1-\gamma)t} \rangle\,.
\label{alphabetatime}
\end{eqnarray}
This result means that the time evolution of the coherent state wavefunction (\ref{psicoherent}) can be obtained replacing $\alpha \to\alpha\,e^{-i\,(1+\gamma)t}$ and $\beta \to \beta\,e^{-i\,(1-\gamma)t}$. 
In order to find the correspondence of classical trajectories and coherent states,
we should identify the eigenvalues $\alpha$, $\beta$ in (\ref{annop}) with the values
$\alpha^-$ and $\beta^-$ of (\ref{abt}).

\begin{figure}[htb]
\centering
\includegraphics[width=0.28\textwidth]{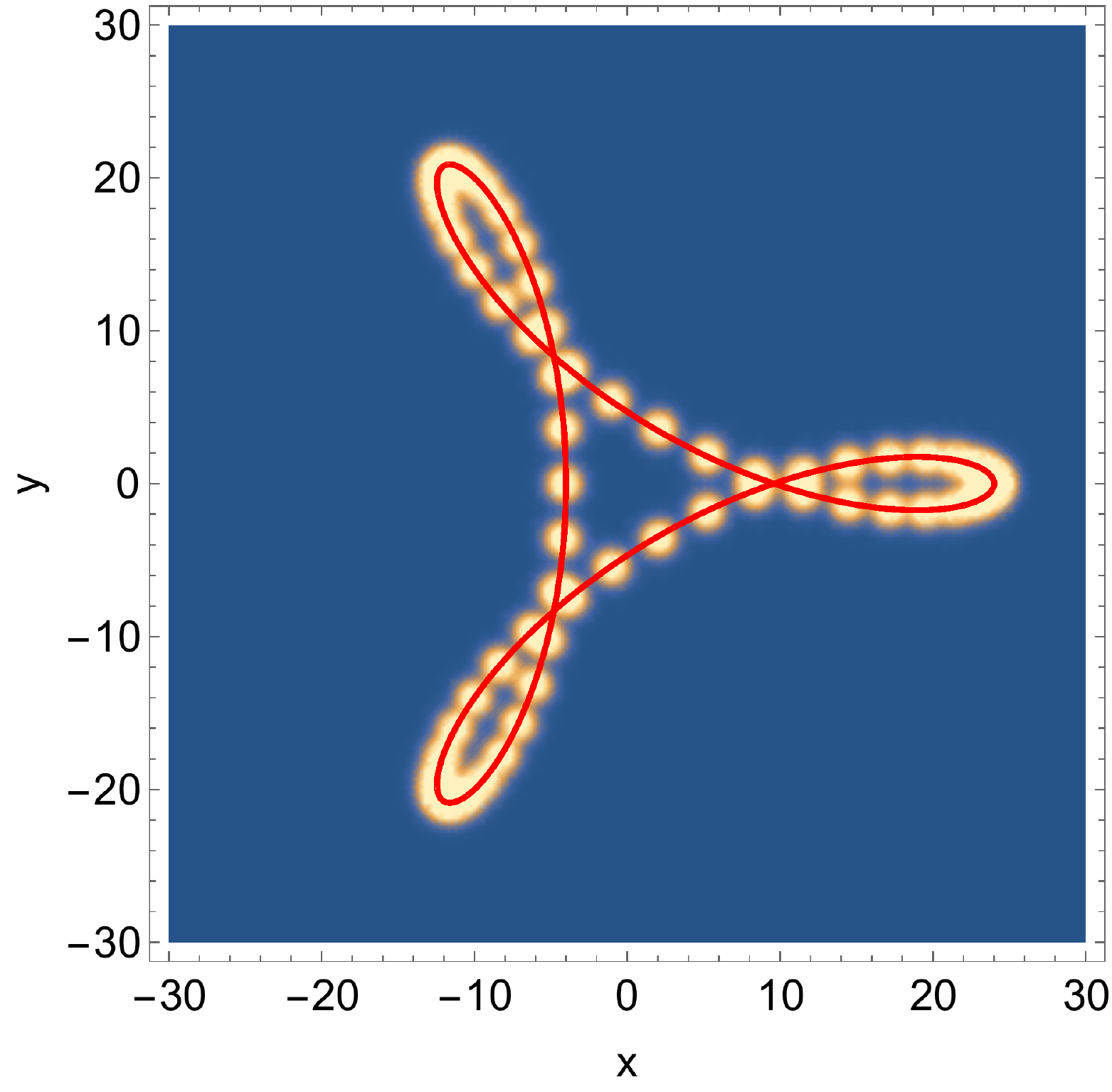}
\qquad 
\includegraphics[width=0.28\textwidth]{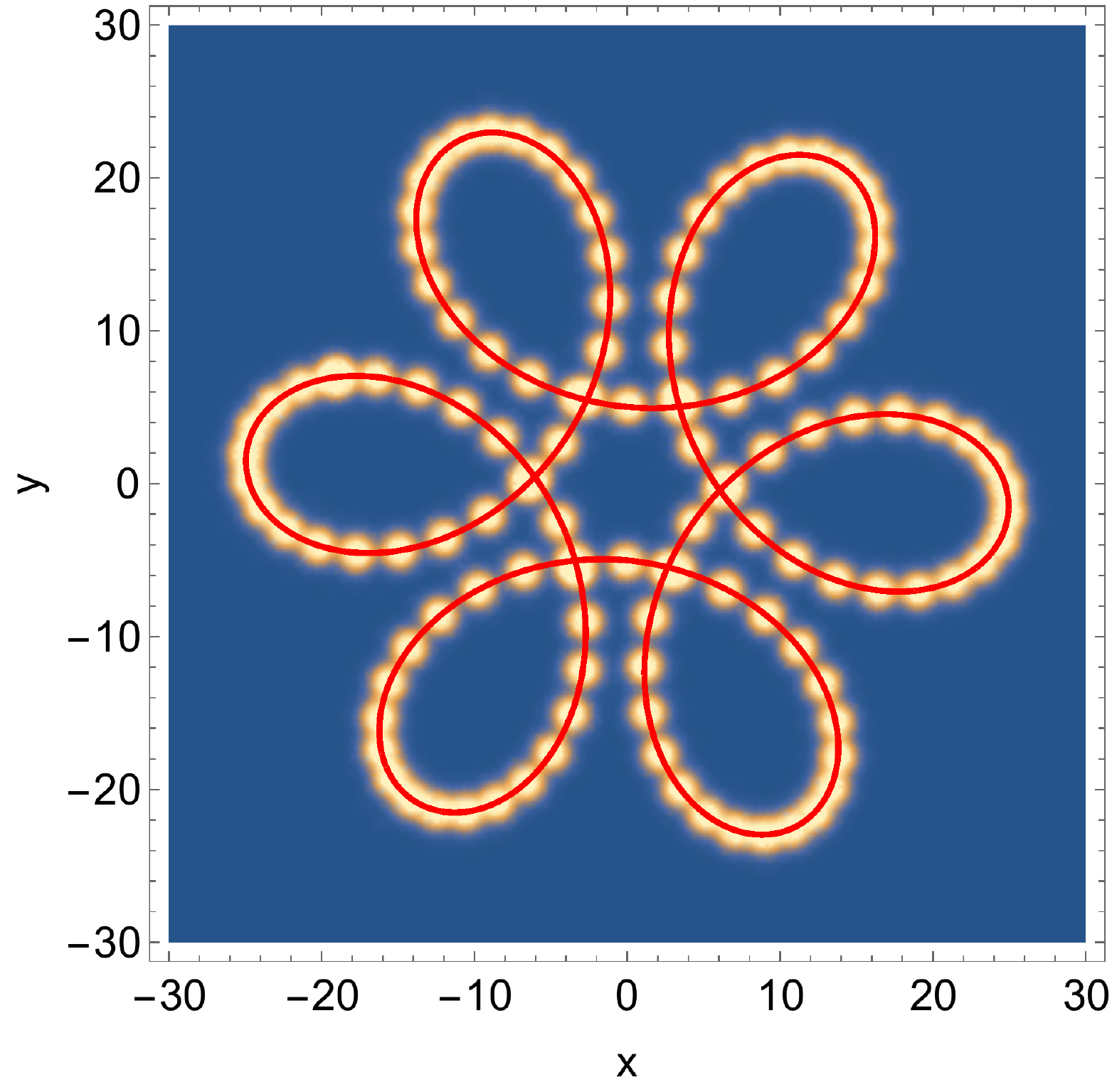}
\qquad 
\includegraphics[width=0.28\textwidth]{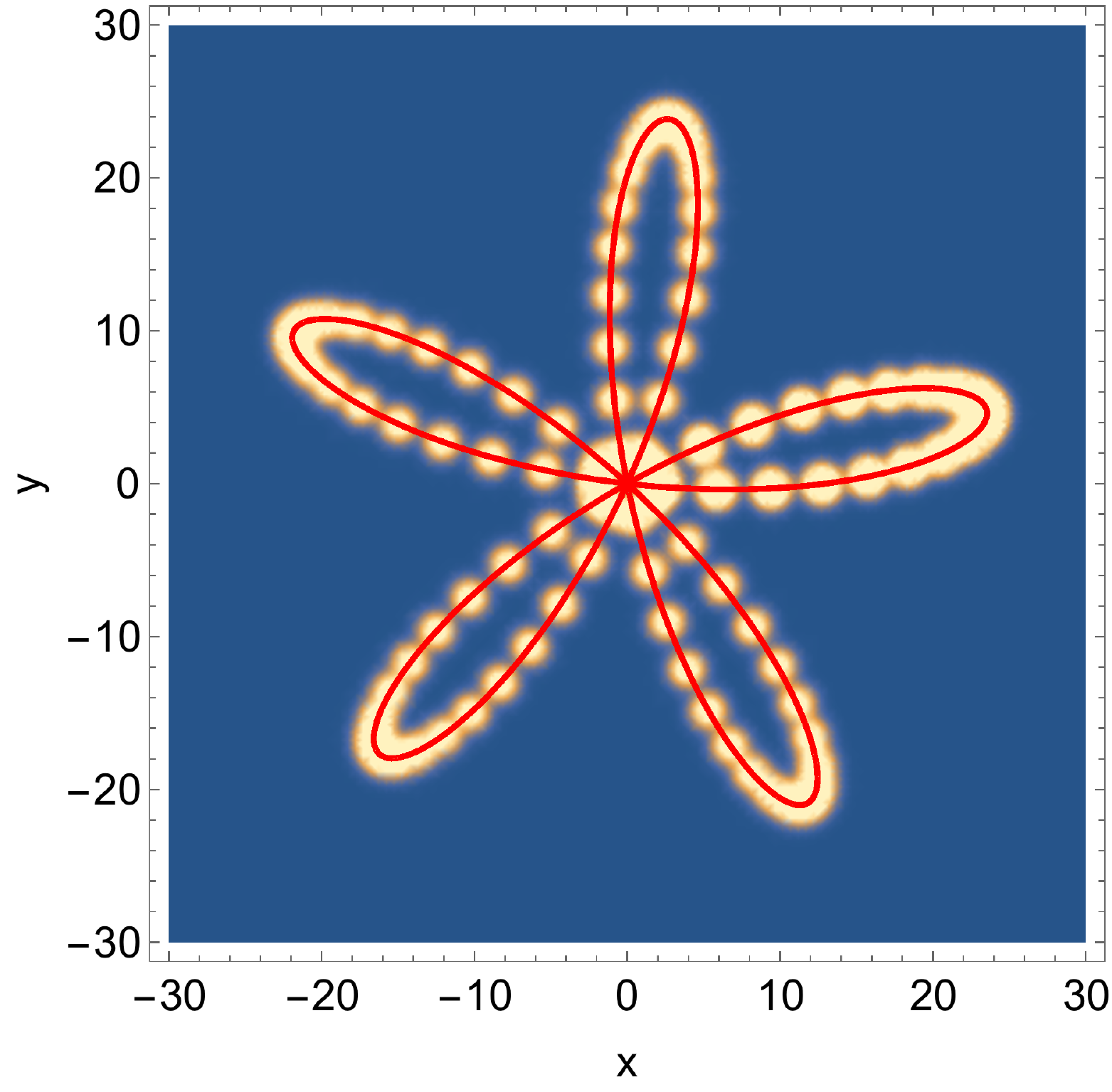}
\caption{\small Plot of the probability density \eqref{quadpsicoherent} of the coherent states and the classical trajectories  (red) for: $\{\gamma=1/3, \alpha_0=10, \beta_0=14, \theta_1=\theta_2=0 \}$ (left),  $\{\gamma=2/3, \alpha_0=10, \beta_0=15, \theta_1=2, \theta_2=3 \}$  (center) and  $\{\gamma=1/5, \alpha_0=12, \beta_0=12, \theta_1=2, \theta_2=1 \}$ (right). 
\label{coherent}}
\end{figure}

In Figure~\ref{coherent}, we plot the probability density of some coherent states and the analogous classical trajectories, which were already considered in Figure~\ref{trajectory}.
From the analysis of both figures it can be seen that the classical trajectories and the expected value of the position coordinates $(x,y)$ for the coherent states are very close.

%%%%%%%%%%%%%%%%%%%%%%%%%

%%%%%%%%%%%%
\sect{Conclusions and remarks}

In this work,  we have systematically studied the symmetry properties of
the FD system, which is characterized by a parameter $\gamma$ relating the frequencies associated to the harmonic oscillator ($\omega_o$) and the magnetic field  ($\omega_c$). We took a different approach from the existing literature, paying attention
to the similarities of symmetries in the quantum and classical frameworks:
the connection of symmetries and degeneracies in the quantum context and 
the relation between constants of motion and transformation of motions with the same energy in the classical case.

Due to its rotational symmetry, the FD system
is separable in polar coordinates  for any value of $\gamma$.  Writing the Hamiltonian in these
coordinates, we have seen that its factorization properties lead to a couple of ladder operators sets: $a^\pm$ and $b^\pm$. Such operators change the energy
as well as the angular momentum of the states and,  from the algebraic point of view, they close a direct sum of two Heisenberg algebras. 
By means of $a^\pm, b^\pm$ we can express the Hamiltonian (\ref{habb}),
which includes the key parameter $\gamma$.
The symmetries of all the FD systems can also be expressed in terms of $a^\pm, b^\pm$.
However, only when $\gamma$ is rational the FD system is superintegrable.

In the particular cases $\gamma=0$ (harmonic oscillator, HO) and $\gamma=\pm1$ (Landau),
the symmetries are  of second order and allow separation in other coordinate
systems (besides polar). In these two cases, the symmetry operators close Lie algebras: $u(2)$
for HO and $os(1)\oplus u(1)$ for Landau.
However, in the other superintegrable rational cases the symmetries are
of higher order and the separation is only possible in polar coordinates. For 
such cases the symmetry algebras are polynomial, and its explicit form was computed.

We have also explained the relation between the symmetries and degeneracy of the energy levels.
The symmetry operators acting on any eigenfunction will generate the whole energy eigenspace.
For the HO the dimension of each eigenspace is $k+1$ ($k=0,1,\dots$), for Landau it is infinite dimensional, and for the FD system is $k+1$, where the number of eigenspaces with
the same degeneracy dimension is $p\times q$.

In the classical FD system, there are ladder functions $\alpha^\pm$ and $\beta^\pm$
corresponding to the quantum operators $a^\pm$ and $b^\pm$. 
Following the same procedure as in the quantum case, we have shown that the FD system
corresponding to rational $\gamma$ values is superintegrable.
We have computed explicitly the Poisson algebra of constants of motion, which are
the classical limit of the corresponding quantum symmetry algebra.

The classical trajectories are directly obtained from the constants of motion. Due to the
superintegrability they are closed, and in this case we have also seen that they 
are $2\pi/(p+q)$ periodic in the polar angle $\varphi$ and they have $2(p+q)$ 
turning points in the variable $\rho$. 
We have studied the action of the constants of motion as generators of symmetry
transformations of the classical motion. In particular, we have been able to give
explicit expressions of the finite action for the HO and Landau. In the general rational FD case the differential equations are nonlinear and a full integration is quite difficult.
However, we have been able to find solutions for some particular cases, showing 
how some motions are transformed by means of finite symmetry transformations.

The connection between the quantum and classical systems is established through coherent states: we have computed explicit expressions of the coherent states in polar coordinates
and we have shown that their evolution follow closely the classical motion trajectories.

In this work, we have restricted to the simplest original FD system, but there are other
models where our considerations also apply. For instance, quantum dots with anisotropic
oscillator confining potentials have been already considered in the literature \cite{Chakraborty94, Chakraborty12}. The superintegrability conditions
can be extended in this case for another characteristic frequency quotient (playing the role of $\gamma$), however the separable coordinates are not polar, but elliptic. In the context of 
paraxial optics, more interaction terms appear in the Hamiltonian  
allowing for a discussion of the conditions to implement superintegrability. Similar properties
have been displayed in some graphics of the coherent states and classical trajectories \cite{Chen10,Chen11}.
Other  variations of the FD model are related with spin Zeeman and Rashba effects, or with the inclusion of electric fields \cite{Chakraborty12}.  Works in these directions are presently in progress.

%%%%%%%%%%%%%%%%%%%%%%%%%%%%%%%%%%%%%%%%%%%%%%%%%%%%%%%
 \section*{Acknowledgments}

This work was partially supported by the Spanish MINECO (MTM2014-57129-C2-1-P) and Junta de Castilla y Le\'on (VA057U16).  \c{S}.~Kuru acknowledges Ankara University and the warm hospitality at Dept.
of Theoretical Physics, Univ. Valladolid, where this work has been done.

%%%%%%%%%%%%%%%%%%%%%%%%%%%%%%%%%%%%%%%%%%%%%%

%%%%%%%%%%%%%%%%%%%%%%%%%%%%%%

\end{document}